# Time-Varying Bidirectional Causal Relationships between Transaction Fees and Economic Activity of Subsystems Utilizing the Ethereum Blockchain Network

**Lennart Ante**[a]

*[a] Blockchain Research Lab*

**Aman Saggu**[ba]

*[b] Business Administration Division, Mahidol University International College, Mahidol University*

*[a] Blockchain Research Lab*

**Abstract:** The Ethereum blockchain network enables transaction processing and smart-contract execution through levies of transaction fees, commonly known as gas fees. This framework mediates economic participation via a market-based mechanism for gas fees, permitting users to offer higher gas fees to expedite processing. Historically, the ensuing gas fee volatility led to critical disequilibria between supply and demand for block space, presenting stakeholder challenges. This study examines the dynamic causal interplay between transaction fees and economic subsystems leveraging the network. By utilizing data related to unique active wallets and transaction volume of each subsystem and applying time-varying Granger causality analysis, we reveal temporal heterogeneity in causal relationships between economic activity and transaction fees across all subsystems. This includes (a) a bidirectional causal feedback loop between cross-blockchain bridge user activity and transaction fees, which diminishes over time, potentially signaling user migration; (b) a bidirectional relationship between centralized cryptocurrency exchange deposit and withdrawal transaction volume and fees, indicative of increased competition for block space; (c) decentralized exchange volumes causally influence fees, while fees causally influence user activity, although this relationship is weakening, potentially due to the diminished significance of decentralized finance; (d) intermittent causal relationships with maximal extractable value bots; (e) fees causally influence non-fungible token transaction volumes; and (f) a highly significant and growing causal influence of transaction fees on stablecoin activity and transaction volumes highlight its prominence. These results inform strategic considerations for stakeholders to more effectively plan, utilize, and advocate for economic activities on Ethereum, enhancing the understanding and optimization of within the rapidly evolving economy.

# 1. Introduction

The Ethereum network is a sophisticated fee market that facilitates economic participation by mediating transaction fees. Firstly, transaction fees incentivize users to employ the network efficiently by ensuring network utilization costs scale with transaction complexity. This approach mitigates the risks of network overload by discouraging users from submitting unnecessary transactions that could impede network performance and increase costs for other users. Secondly, transaction fees contribute to the economic sustainability of the Ethereum network by compensating miners (prior to 15 September 2022) or validators (after 15 September 2022) for their network operation services. This incentivization model rewards their contribution to the network's security and stability, strengthening overall network performance and resilience. Thirdly, transaction fees foster decentralization of the Ethereum network by leveling the playing field for all users, thereby removing barriers to entry for new users or developers.

As a decentralized platform, the Ethereum network enables users to create and deploy smart contracts, which are programmatic transaction protocols that automatically execute the terms of an agreement between buyers and sellers. These stipulations are directly inscribed into the code, making them more transparent and secure. To compensate for the computational resources consumed by the network during transaction execution, users pay transaction fees (Szabo 1994). Since its genesis in 2015, the Ethereum network has engendered the emergence of an array of decentralized economic systems, each reliant upon the underlying transaction costs and network scalability in unique ways. Transactions on the Ethereum network are processed by validating new blocks, on average, every 12 seconds. Each block consolidates a variable number of initiated transactions, which are subsequently confirmed by the network.[1] In 2022, the network processed an average of approximately 1.2 million transactions daily (Etherscan 2022). However, network scalability is inherently constrained, as only a specific number of transactions can be validated within each 12-second interval. Historically, this constraint has caused critical disequilibria between supply and demand.

The rapid adoption of high-profile blockchain-based applications has given rise to numerous challenges, including network congestion and increased transaction costs. A salient example is the late 2017 phenomenon of CryptoKitties, a game employing blockchain technology to enable the buying, collecting, breeding, and selling of digital cats in the form of non-fungible tokens (NFTs). The unprecedented demand for these digital collectibles culminated in a record-high volume of transactions on the network, consequently inducing severe congestion and a surge in transaction costs (BBC 2017; Ante 2022). Another manifestation of this phenomenon can be observed in digital token sales, conducted by blockchain enterprises, which encountered extraordinary investor demand, leading to the so-called gas wars (Spain et al. 2020). The term gas refers to the internal metering variable of the Ethereum network, which necessitates the payment of gas for every transaction execution.[2] Heightened demand for digital token sales in 2017 and 2018 prompted a spike in gas prices, rendering transactions on the Ethereum network increasingly costly. Furthermore, instances of fundamentally high network utilization also had a detrimental impact on the efficiency of individual applications. In 2021, users faced abnormally high transaction fees when exchanging Ethereum-based tokens on decentralized exchanges (DEXs) such as Uniswap. These fees occasionally surpassed several hundred dollars, rendering transactions prohibitively costly (Etherscan 2021). Such high transaction costs, coupled with the limitations imposed by network congestion, rendered the Ethereum network less accessible and efficient, prompting individual users, protocols, and applications to suspend their activity or migrate to other blockchains (Axie 2021).

Numerous studies have explored the Ethereum fee market from technical or modeling perspectives (Reijsbergen et al. 2021; Laurent et al. 2022; Azevedo Sousa et al. 2021; Werner et al. 2020; Bouraga 2020), often less focusing on the economic foundations of the Ethereum ecosystem.[3] However, it is crucial to acknowledge the instrumental role of economic incentives and developments in shaping Ethereum's trajectory and fee market, as evidenced by phenomena such as CryptoKitties, token sales, and the emergence of DEXs, which profoundly impacted Ethereum's evolution. Incorporating an economic perspective into the analysis of Ethereum's fee market is essential for obtaining a comprehensive understanding of the ecosystem. While prior studies have examined individual systems and unanticipated events in specific systems (Faqir-Rhazoui et al. 2021; Spain et al. 2020), there remains a lack of comprehensive analyses encompassing multiple economic systems and their potential interdependencies within the Ethereum fee market context. This research gap underscores the necessity for a thorough investigation of the economic environment in which fees function, to comprehend their relevance. Such an analysis would constitute a valuable reference for decision-making in projects intending to utilize Ethereum as an infrastructure. Additionally, from a stakeholder-oriented perspective (Freeman 1984), it is important to assess the extent to which projects can tolerate comparatively less favorable fees, in comparison to layer-2 solutions such as Arbitrum or Optimism, given the potential network and spillover effects that may derive from other markets or assets on Ethereum. This would allow stakeholders to make informed decisions and adapt their strategies to the dynamic Ethereum fee market.

Our research aims to examine the activity of economic systems, classified based on their direct or indirect dependency on the Ethereum network, and evaluate their influence on transaction fees. For example, maximal extractable value (MEV) (Daian et al. 2020) may only be significant as long as (uninformed) market participants participate in the Ethereum network, as observed with DEXs and meme coins (Li and Yang 2022; Xia et al. 2021). In light of this context, we pose the research question: what significant economic activities transpire within the Ethereum network, and how do they affect transaction fees? This inquiry is crucial for understanding the broader economic implications of blockchain technology and its impact on various stakeholders within this rapidly evolving digital landscape.

To identify economic systems, we undertake an exploratory multivocal analysis of Ethereum data platforms, informed by the existing scientific literature on Ethereum. Following the identification of these systems, we delve into the central inquiry of this study: uncovering the causal relationship between individual systems and fees within the Ethereum network.[4] We employ two proxies as measures for each system's economic activity: (a) the number of users specific to a particular system and (b) the on-chain economic value transferred (in USD) for that system. This approach ensures that high-usage solutions with low volumes receive due attention, while also considering applications with relatively few users but significant transaction volumes. In essence, the two variables serve as complementary metrics for evaluating economic participation and relevance within the network.

Given the temporal sensitivity of causality (Shi et al. 2018; Ren et al. 2023), this study utilizes the time-varying Granger causality methodology to discern the relationship between fees, blockchain users (represented by unique active wallets), and economic volume. This econometric methodology, devoid of pre-processing steps, is adept at identifying the inception and cessation dates of causal episodes. A comprehensive understanding of the bidirectional causality between these variables enables the revelation of the temporal trajectory of the causal relationship. Employing time-varying Granger causality for the identification of causal relationships further facilitates a plethora of research opportunities, encompassing an in-depth analysis of the temporality and directionality of relationships (i.e., whether they are positive or negative).

This study makes several contributions to the literature, aligning with broader research themes in the fields of blockchain technology, network economics, and digital finance. Firstly, it reveals that transaction fees play an instrumental role in the functioning of a majority of the examined systems. The findings suggest that these fees are not merely a technical feature but a critical economic lever affecting the viability and dynamics of blockchain-based (eco-)systems. This aspect of the research contributes to the growing body of work examining the microstructure of blockchain networks and the economic behavior of digital asset markets. Additionally, the activity and volume of numerous systems exert a considerable impact on network fees. It is worth noting that the causal relationships among variables exhibit dynamic properties and are subject to temporal fluctuations. This variability can be attributed to a combination of factors, including the microstructure of endogenous factors within the Ethereum ecosystem, as well as exogenous factors that are outside the scope of this investigation. This highlights the importance of considering broader market dynamics and technological developments in blockchain research. Secondly, the results indicate that the number of unique active wallets influences the average fees in the network, and fees play a vital role in user migration. Consequently, fees ought to be judiciously managed by Ethereum network stakeholders. The causal relationship between fees, users, and the volume of deposits and withdrawals on cryptocurrency exchanges underscores the presence of a time-varying feedback loop within the Ethereum network's functioning. This adds to the understanding of market mechanisms in digital finance and platforms, particularly in the context of decentralized financial structures.

Thirdly, the investigation of the causal spillover between DEX volume, DEX users, and network fees emphasizes the necessity for DEXes to strike a balance between attracting more users and maintaining manageable fees. The diminishing significance of fees over time, for instance, for DEXes, implies that fees' impact on users may vary depending on the economic context, signifying the relevance of MEV for the Ethereum network at certain times. This aspect of the study aligns with broader research themes in blockchain scalability, user-centric design in DeFi, and the economics of transaction fees in distributed ledger technologies. In conclusion, these findings provide valuable insights into the operation of these economic systems on Ethereum. They not only enhance our understanding of the interdependencies within the ecosystem but also stress the importance of ongoing research into the multifaceted relationship between transaction fees and network activity. Such research is vital for developing more efficient, user-friendly, and economically viable blockchain platforms, which are key goals in the broader field of digital finance and blockchain technology development.

This article proceeds as follows. Section 2 provides a conceptual background on transaction costs from a theoretical perspective (Section 2.1) and with a focus on the Ethereum network (Section 2.2). In Section 3, the process of identifying economic systems on the blockchain network (Section 3.1), data collection and descriptive considerations (Section 3.2), and the empirical approach are described (Section 3.3). Section 4 consists of the results of the time-varying Granger causality. Section 5 discusses the results and concludes.

## 2. Conceptual Background

### *2.1. Transaction Cost Theory*

Transaction cost theory, an economic concept that focuses on the costs associated with conducting transactions, posits that these costs can significantly impact the efficiency of resource allocation, regardless of the distribution of property rights. The theory is based on Coase (Coase 1937), was initially proposed by Williamson (Williamson 1979) in the late 1970s, and was subsequently developed and refined for specific use cases (Masten et al. 1991; Hennart 1982). In the context of Ethereum, the theory can help understand the role of transaction costs, such as gas fees, in allocating resources within the network. There are significant transaction costs in the Ethereum ecosystem, some of which include:

- *Information costs*: Imperfect information, price volatility, and complexity contribute to information transaction costs in the Ethereum network. Users may lack accurate information about gas prices, leading to overpaying or underpaying, thereby contributing to resource allocation inefficiencies. Gas price volatility and the intricacies of understanding and calculating gas fees further complicate decision-making for users, especially for those without technical expertise in blockchain technology (Abel et al. 2013; Holmstrom and Milgrom 1991; Arrow 1974).
- *Bargaining costs*: The Ethereum network handles a vast number of transactions daily, making individual bargaining for each transaction time-consuming and impractical. Ethereum is designed to maintain pseudonymity, and direct negotiation of fees could jeopardize this anonymity. Additionally, the dynamic nature of gas prices and fluctuating demand for transaction fee negotiation are challenging and unrealistic (Milgrom and Roberts 1990; Hart and Moore 1990; Grossman and Hart 1986).
- *Enforcement costs*: In Ethereum's trustless, decentralized environment, enforcing agreed-upon fees and transaction inclusion can be challenging without a centralized authority. Resolving disputes related to transaction fees or performance is difficult and costly. Aligning incentives for users and miners or validators is crucial and can be achieved through well-designed economic mechanisms and consensus algorithms (Dyer and Singh 1998; Zaheer and Venkatraman 1995).

The Ethereum network employs a multifaceted approach to addressing information, bargaining, and enforcement costs. To mitigate information costs, gas price estimators have been developed, to provide real-time estimates to aid users in selecting suitable fees. Additionally, Ethereum's market-based mechanism for determining gas prices allows users to specify fees, while miners (before September 2022) or validators (after September 2022) opt for transactions based on the fees offered, approximating efficient resource allocation. Ongoing protocol upgrades, such as Ethereum 2.0, strive to enhance the predictability and user-friendliness of the gas fee market. As for enforcement costs, the network leverages smart contracts to automatically enforce agreements, thereby minimizing manual enforcement and dispute resolution. Ethereum relies on consensus algorithms, such as PoW (before September 2022) and PoS (after September 2022), to maintain blockchain integrity, align incentives, and ensure the inclusion of transactions. Furthermore, fee markets create incentives for miners to prioritize transactions with higher fees and for users to pay appropriate fees to facilitate prompt transaction processing.

### *2.2. Ethereum Transaction Fees*

Ethereum is a blockchain platform that uses gas to execute transactions and host smart contracts (Buterin 2014). The amount of gas used for a transaction is determined by the computing power required for that specific contract. The transaction fee for each execution is based on a free-market system, where the issuer decides how much they are willing to pay for each unit of gas. Miners or validators determine which transactions are included in blocks, e.g., the ones with the highest fees. However, the flexibility and complexity of this system present challenges for developers, maintainers, and users of blockchain-powered applications (Khan et al. 2022). Based on an analysis of the gas usage of Ethereum transactions between October 2017 and February 2019, Zarir et al. (2021) find that the majority of miners prioritized transactions based on gas prices alone. Further, the authors show that 25% of functions with at least 10 transactions have unstable gas usage and suggest that developers can use prediction models to make more informed decisions on gas prices. Another study finds that increasing gas prices does not significantly reduce the end-to-end latency of Ethereum within a certain range of prices (Zhang et al. 2021).

Donmez and Karaivanov (2022) investigate the economic factors that influence transaction fees within the Ethereum blockchain. Through the use of queueing theory and empirical analysis, they show that changes in service demand have a significant impact on fees. Specifically, when blockchain utilization is high, per-unit fees tend to increase on average, with a particularly strong nonlinear effect observed above 90% utilization. Additionally, they identify that the type of transaction also plays a crucial role in determining the gas price, with a larger proportion of regular transactions (i.e., direct transfers between users) being associated with higher gas prices. Another study that relies on data from 2018 shows that the number of pending transactions and the number of miners have the greatest influence on Ethereum transaction fees when compared to other factors (Pierro and Rocha 2019).

An examination of the impact of transaction fee prices on user activity in blockchain-enabled decentralized systems, specifically focusing on Decentralized Autonomous Organizations (DAOs) found that there is only a minor influence of fee (gas) prices on user activity, which is anomalous in a self-regulated market (Faqir-Rhazoui et al. 2021). Focusing on the context and impact of the competitive environment among buyers vying for a limited supply of tokens offered in initial coin offerings (ICOs), i.e., token sales, Spain et al. (2020) indicate that while buyers incentivize miners to prioritize their transactions during ICOs, the latency of these transactions is primarily determined by the levels of supply and demand in the network.

Through the collection of information on 7.2 million Ethereum transactions, Azevedo Azevedo Sousa et al. (2021) correlate the pending time of transactions to several fee-related features and evaluate different ranges of values for these features, including default and unusual values adopted by users and clusters of users with similar behaviors. The results of the empirical analysis provide strong evidence that there is no clear correlation between fee-related features and the pending time of transactions. Therefore, the authors conclude that transaction features, including gas and gas prices defined by users, cannot determine the pending time of transactions on the Ethereum platform. A plethora of other studies analyze, evaluate or forecast gas prices on Ethereum, i.e., how fees should ideally be set for the initiation of transactions, and find multiple ways for optimization or identify inefficiencies (Feng et al. 2023; Bouraga 2020; Antonio Pierro et al. 2020; Pierro et al. 2022; Mars et al. 2021; Chunmiao Li et al. 2020; Liu et al. 2020; Laurent et al. 2022; Werner et al. 2020).

The Ethereum Improvement Proposal (EIP) 1559 was implemented to improve the transaction fee market on Ethereum. The update uses an algorithmic rule with a constant learning rate to estimate a base fee, which reflects current network conditions (Buterin et al. 2019; Yulin Liu et al. 2022). However, research on on-chain data from the period after its launch suggests that EIP-1559 has led to intense, chaotic oscillations in block sizes and slow adjustments during periods of high demand. These phenomena result in unwanted variability in mining rewards (Leonardos et al. 2021). To address this issue, Reijsbergen et al. (2021) propose an alternative base fee adjustment rule that utilizes an additive increase multiplicative decrease (AIMD) update scheme and provides simulations showing that the approach outperforms EIP-1559. Also referring to limitations of mechanisms such as EIP-1559, Chunmiao Li (2021) proposes a dynamic posted-price mechanism, which uses not only block utilization but also observable bids from past blocks to compute a posted-price for subsequent blocks. The goal of this mechanism is to reduce price volatility.

In conclusion, it can be argued that there is a significant amount of ongoing research in the area of transaction costs within the Ethereum network. However, the majority of this research is concentrated within the field of computer science, specifically focusing on identifying optimization opportunities and developing novel approaches. In contrast, there exists comparatively limited research from an economic or transaction cost theory perspective or research that specifically examines specific economic systems or phenomena. This serves as motivation for the above-mentioned research question and methodological approach outlined in the subsequent sections.

# 3. Methods and Data

## *3.1. Ethereum Transaction Fees*

To identify the economic systems of the Ethereum network for empirical analysis, a rigorous, multivocal exploratory approach is employed. This approach consists of two primary steps: (1) an exploratory analysis of non-academic data platforms in the field of Ethereum, focusing on addresses or address bundles associated with numerous transactions and fees; and (2) a validation of the identified systems through bibliometric analyses of the respective topics, determining the relevance of the systems as recognized by academic literature.

For the initial step, Ethereum and blockchain data providers such as Etherscan (Etherscan 2023b), Dune Analytics (Dune Analytics 2023), Glassnode (Glassnode 2020), and Flipside Crypto (Flipside Crypto 2023) are utilized to analyze the Ethereum network's status at regular one-week intervals between 1 August and 1 November 2022. This analysis aims at identifying and documenting relevant markets, wallets, contracts, and estimated transaction costs, resulting in a preliminary overview of significant economic systems. For instance, the Ethereum addresses of stablecoins Tether (USDT) and USD Coin (USDC) were found to be associated with a high number of transactions, thus indicating their substantial economic significance. Consequently, the "stablecoins" system was identified and validated based on the abundant academic literature on the subject. In contrast, the Ethereum Name Service (ENS), an open naming system based on the Ethereum blockchain, represents an atypical case. Although on-chain data suggest its relevance as a "system", academic validation is currently lacking. As such, ENS may represent a promising area for future research.

Table 1 presents a summary of the six identified systems examined in this study, accompanied by pertinent academic references for validation purposes. These systems include (a) Bridges, (b) Centralized Exchanges (CEXs), (c) Decentralized Exchanges (DEXs), (d) Maximal Extractable Value (MEV) bots, (e) Non-Fungible Tokens (NFTs), and (f) Stablecoins. It is important to note that this selection is not intended to be a comprehensive compilation of every economic system on the Ethereum network, but rather a representative sample of systems that are relevant to the research question. Given the challenge of comprehensively identifying, extracting data from, and analyzing all six economic systems in question, we utilize proxy variables to approximate the economic activity of individual systems. Specifically, we employed Etherscan's Ethereum address labeling service, which facilitates the identification and classification of prominent actors on the Ethereum blockchain through the assignment of account labels. For instance, a roster of all recognized bridges can be accessed through the account handle, "bridge" (Etherscan 2023a). This approach was replicated for all identified economic systems, yielding a range of three to five addresses or contracts as proxy variables per system, which are presented in Table 1. These addresses were then used to extract data on volumes and distinct active users, which were aggregated to compute system-specific data.

**Table 1. Economic systems and proxy addresses on the Ethereum blockchain.**

| System | Description | Name | Address | Creation Date/ First Transaction | Number of Transactions |
|---|---|---|---|---|---|
| Bridges | Blockchain protocols or platforms that allow for interoperability between different blockchain networks (Lan et al. 2021; Teutsch et al. 2019; Zhang et al. 2022; Lee et al. 2022; Xie et al. 2022; Yiying Liu et al. 2022; Qasse et al. 2019; Belchior et al. 2022; Stone 2021; Hardjono 2021). | Axie Infinity: Ronin Bridge | 0x1a2a1c938ce3ec39b6d47113c7955baa9dd454f2 | 25 January 2021 | 3,090,670 |
| | | zkSync | 0xabea9132b05a70803a4e85094fd0e1800777fbef | 15 June 2020 | 825,134 |
| | | Hop Protocol | 0xb8901acb165ed027e32754e0ffe830802919727f | 1 October 2021 | 497,580 |
| | | Immutable X: Bridge | 0x5fdcca53617f4d2b9134b29090c87d01058e27e9 | 10 March 2021 | 384,515 |
| | | Optimism: Gateway | 0x99c9fc46f92e8a1c0dec1b1747d010903e884be1 | 22 June 2021 | 300,528 |
| CEX | Deposits and withdrawals from wallets of centralized crypto asset exchanges (Ante et al. 2021a; le Pennec et al. 2021; Brandvold et al. 2015; Makarov and Schoar 2020; Ante 2020; Bianchi et al. 2022; Petukhina et al. 2021). | Binance Hot Wallet A | 0x3f5ce5fbfe3e9af3971dd833d26ba9b5c936f0be | 4 August 2017 | 17,017,383 |
| | | Binance Hot Wallet B | 0x28c6c06298d514db089934071355e5743bf21d60 | 22 April 2021 | 11,507,057 |
| | | Bittrex Wallet | 0xfbb1b73c4f0bda4f67dca266ce6ef42f520fbb98 | 10 August 2015 | 11,492,410 |
| | | Coinbase Wallet A | 0x3cd751e6b0078be393132286c442345e5dc49699 | 27 April 2021 | 9,852,269 |
| | | Coinbase Wallet B | 0xb5d85cbf7cb3ee0d56b3bb207d5fc4b82f43f511 | 27 April 2021 | 9,351,971 |
| DEX | Decentralized exchanges (DEX), which allow for peer-to-peer trading of crypto assets (Lan et al. 2021; Teutsch et al. 2019; Zhang et al. 2022; Lee et al. 2022; Xie et al. 2022; Yiying Liu et al. 2022; Qasse et al. 2019; Belchior et al. 2022; Stone 2021; Hardjono 2021). | SushiSwap Router | 0xd9e1ce17f2641f24ae83637ab66a2cca9c378b9f | 4 September 2020 | 4,131,024 |
| | | Uniswap v2 Router | 0x7a250d5630b4cf539739df2c5dacb4c659f2488d | 5 June 2020 | 58,660,014 |
| | | Uniswap v3 Router | 0xe592427a0aece92de3edee1f18e0157c05861564 | 4 May 2021 | 5,673,190 |
| MEV | Bots that exploit market inefficiencies to extract profit, known as miner extractable value or maximal extractable value (MEV) (Daian et al. 2020; Qin and Gervais 2021; Zhou et al. 2021; Churiwala and Krishnamachari 2022; Obadia et al. 2021; Kulkarni et al. 2022; Malkhi and Szalachowski 2022; Weintraub et al. 2022; Bartoletti et al. 2022). | MEV Bot 1 | 0xa57bd00134b2850b2a1c55860c9e9ea100f26dd6cf | 26 March 2019 | 3,641,491 |
| | | MEV Bot 2 | 0x0000000000007f150bd6f54c40a34d7c3d5e9f56 | 23 October 2020 | 2,327,098 |
| | | MEV Bot 3 | 0x860bd2dba9cd475a61e6d1b45e16c365f6d78f66 | 11 February 2020 | 2,175,487 |
| | | MEV Bot 4 | 0x000000000000006f6502b7f2bbac8c30a3f167e9a | 1 May 2020 | 1,438,193 |
| | | MEV Bot 5 | 0x4cb18386e5d1f34dc6eea834bf3534a970a23f8e7 | 6 February 2021 | 732,871 |
| NFTs | Non-fungible tokens, which are unique digital assets that can represent ownership of things like artwork or collectibles (Daian et al. 2020; Qin and Gervais 2021; Zhou et al. 2021; Churiwala and Krishnamachari 2022; Obadia et al. 2021; Kulkarni et al. 2022; Malkhi and Szalachowski 2022; Weintraub et al. 2022; Bartoletti et al. 2022). | Azuki | 0xed5af388653567af2f388e6224dc7c4b3241c544 | 20 January 2022 | 87,238 |
| | | Bored Ape Yacht Club | 0xbc4ca0eda7647a8ab7c2061c2e118a18a936f13d | 22 April 2021 | 141,249 |
| | | CloneX | 0x49cf6f5d44e70224e2e23fdcdd2c053f30ada28b | 12 December 2021 | 100,589 |
| | | Mutant Ape Yacht Club | 0x60e4d786628fea6478f785a6d7e704777c86a7c6 | 28 August 2021 | 121,620 |
| | | CryptoPunks | 0xb47e3cd837ddf8e4c57f05d70ab865de6e193bbb | 22 June 2017 | 50,630 |
| Stablecoins | Crypto assets that are pegged to the value of a specific asset, such as the USD dollar, in order to reduce volatility in their value (Fiedler and Ante 2023; Hoang and Baur 2021; Briola et al. 2022; Ante et al. 2021b; 2021a; Grobys et al. 2021; Moin et al. 2020; Saggu 2022; Griffin and Shams 2020; Wang et al. 2020). | BUSD | 0x4fabb145d64652a948d72533023f6e7a623c7c53 | 5 September 2019 | 1,740,628 |
| | | AI | 0x6b175474e89094c44da98b954ee-deac495271d0f | 13 November 2019 | 16,642,305 |
| | | FRAX | 0x853d955acef822db058eb8505911ed77f175b99e | 16 December 2020 | 575,751 |
| | | USDC | 0xa0b86991c6218b36c1d19d4a2e9eb0ce3606eb48 | 3 August 2018 | 59,242,016 |
| | | USDT | 0xdac17f958d2ee523a2206206994597c13d831ec7 | 28 November 2017 | 174,406,655 |

*Table 1 provides a comprehensive overview of six unique economic systems operating on the Ethereum blockchain, along with their corresponding proxy addresses. The systems analyzed encompass Bridges, Centralized Exchanges (CEXs), Decentralized Exchanges (DEXs), Maximal Extractable Value (MEV) bots, Non-Fungible Tokens (NFTs), and Stablecoins, with data sourced from Etherscan. The table offers a succinct description of each system, the precise proxy address, the inception date or initial transaction, and the total number of transactions linked to the address. The table features a concise description of each system, the specific proxy address, the creation date or first transaction, and the aggregate number of transactions associated with the address. All transaction data were collected on 28 January 2022.*

### *3.2. Data*

#### 3.2.1. Transaction Fees on the Ethereum Blockchain

We utilize Glassnode (Glassnode 2020) as our principal data source to acquire comprehensive time series data on transaction costs (fees) within the Ethereum network at daily intervals. The dataset covers the period from 1 July 2020 to 14 November 2022 and encompasses mean transaction fees in USD. Across all transaction types, the mean daily value of fees amounts to USD 13.18, indicating that they are conceivably "too high" for a plethora of applications that necessitate substantial transaction throughput.[5] To illustrate the behavior of fees over time, we depict the mean fees in USD in both logarithmic form and as a first difference in Figure 1. Notably, we find no discernible trend in the logarithmic form of the variable.

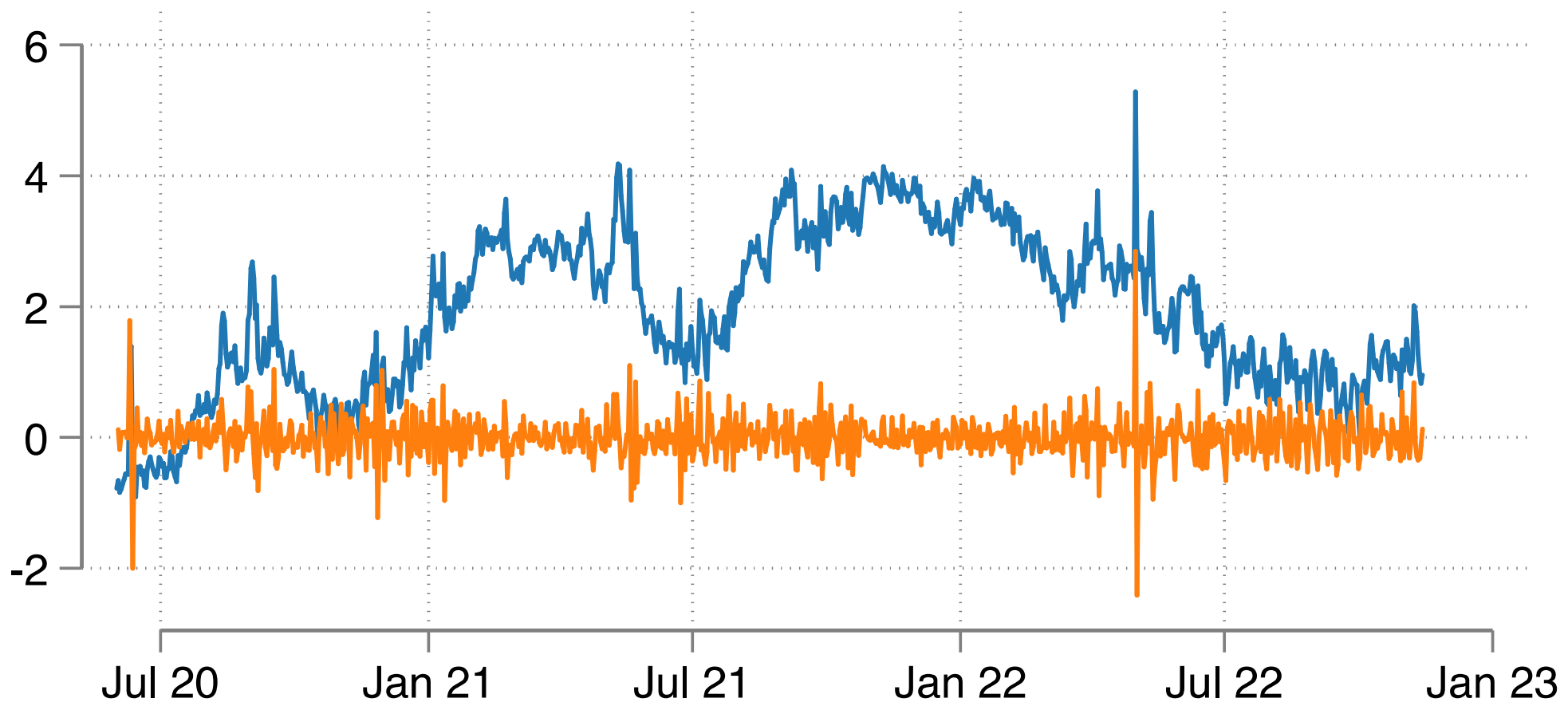

***Figure 1.*** *Transaction fees on the Ethereum blockchain. Figure 1 illustrates the transaction fees on the Ethereum blockchain using data obtained from Glassnode at daily intervals. The dataset comprises mean transaction fees in USD from 1 July 2020 to 14 November 2022. The blue curve represents the logged transaction fees, while the orange curve displays the first difference in the transaction fees.*

To investigate the time series properties of the transaction fee variable, we apply the ADFmax unit root test because it accounts for both forward and reverse realizations of the variable under scrutiny (Leybourne 1995; Otero and Baum 2018). We incorporate a constant and trend in the test regressions. The selection of lag order is informed by the Akaike Information Criterion (AIC), the Schwarz Information Criterion (SIC), and a General-to-Specific algorithm (Hall 1994; Campbell and Perron 1991) with a 5% significance level. The results of the test are displayed in Table 2. The unit root tests indicate that fees are optimally represented in first differences, as the hypothesis of a unit root cannot be rejected for the log-transformed series.[6]

**Table 2. ADFmax test results for time series properties of Ethereum transaction fees in USD.**

| | **Log-Transformed** | | | **First-Differenced** | | |
|---|---|---|---|---|---|---|
| | **Lags** | **ADFmax** | ***p*-Value** | **Lags** | **ADFmax** | ***p*-Value** |
| AIC | 6 | −1.018 | 0.536 | 7 | −12.233 *** | 0.000 |
| SIC | 5 | −1.196 | 0.444 | 4 | −19.770 *** | 0.000 |
| $GTS_{05}$ | 6 | −1.018 | 0.537 | 7 | −12.233 *** | 0.000 |

*Table 2 displays the results of an ADFmax test, incorporating a constant and trend for both log-transformed and first-differenced transaction fees on the Ethereum blockchain. The data are sourced from Glassnode, with daily intervals covering the period between 1 July 2020 and 14 November 2022. The untransformed dataset considers mean transaction fees in USD. Significance levels are indicated by *, **, and *** for 10%, 5%, and 1%, respectively.*

#### 3.2.2. Underlying Economic Systems on the Ethereum Blockchain

We utilized Flipside Crypto (Flipside Crypto 2023) as a data source to collect daily time series data on the economic systems within the Ethereum network. To achieve this, we developed and implemented custom application programming interfaces (APIs) for each distinct contract or address on the Ethereum blockchain defined in Table 1. These APIs allowed us to determine the number of unique active wallets and the corresponding volume, measured in USD, associated with transactions. The term "active" denotes a wallet, specifically a blockchain address, which partakes directly in a successful transaction. Conversely, the term "unique" signifies that addresses are not enumerated multiple times, thereby precluding redundancy. Consequently, the metric of unique active wallets can serve as a surrogate measure for authentic singular users within the blockchain ecosystem. It is imperative to note, however, that the determination of whether a lone individual exercises control over multiple unique addresses remains unfeasible.

Figure 2 presents time series plots for the six identified economic systems, showcasing the transacted volume in USD and unique active wallets, both log-transformed. In the case of the (a) bridge system, a considerable upsurge in volume is observed, commencing in July 2021 and reaching its zenith towards the end of 2021 and mid-2022. On average, the daily transfer volume via the bridges is slightly above USD 23 million, with a highly skewed distribution (standard deviation = USD 77 million; maximum = USD 1.5 billion). The number of unique active wallets also experiences substantial growth, escalating from less than 1000 to almost 20,000 wallets per day during mid-2021. Notwithstanding the subsequent decline in active wallets, it is essential to note that the activity levels remain markedly elevated compared to earlier periods, with a peak value of 29,793 unique active wallets recorded in June 2022. On average, 2757 unique wallets per day interact with the bridges employed for system calculation, highlighting the system's economic importance.

An analysis of (b) CEX deposit and withdrawal activity in Figure 2 reveals a striking growth pattern in volume up to mid-2021, succeeded by a precipitous decline over several months and a brief resurgence towards the end of 2022. Subsequently, a sustained decrease in average volume emerges. The daily mean economic volume stands at USD 717 million, accompanied by a standard deviation of USD 686 million, indicating considerable variability in the data. The number of distinct active wallets mirrors the volume fluctuations, averaging 25,205 per day and peaking at 590,000 on a single day—an exceptional outlier. Notably, sporadic short-term user surges in 2022 are ephemeral aberrations that do not alter the overarching downward trajectory, reflecting a potential shift in user behavior and market dynamics within the centralized exchange ecosystem.

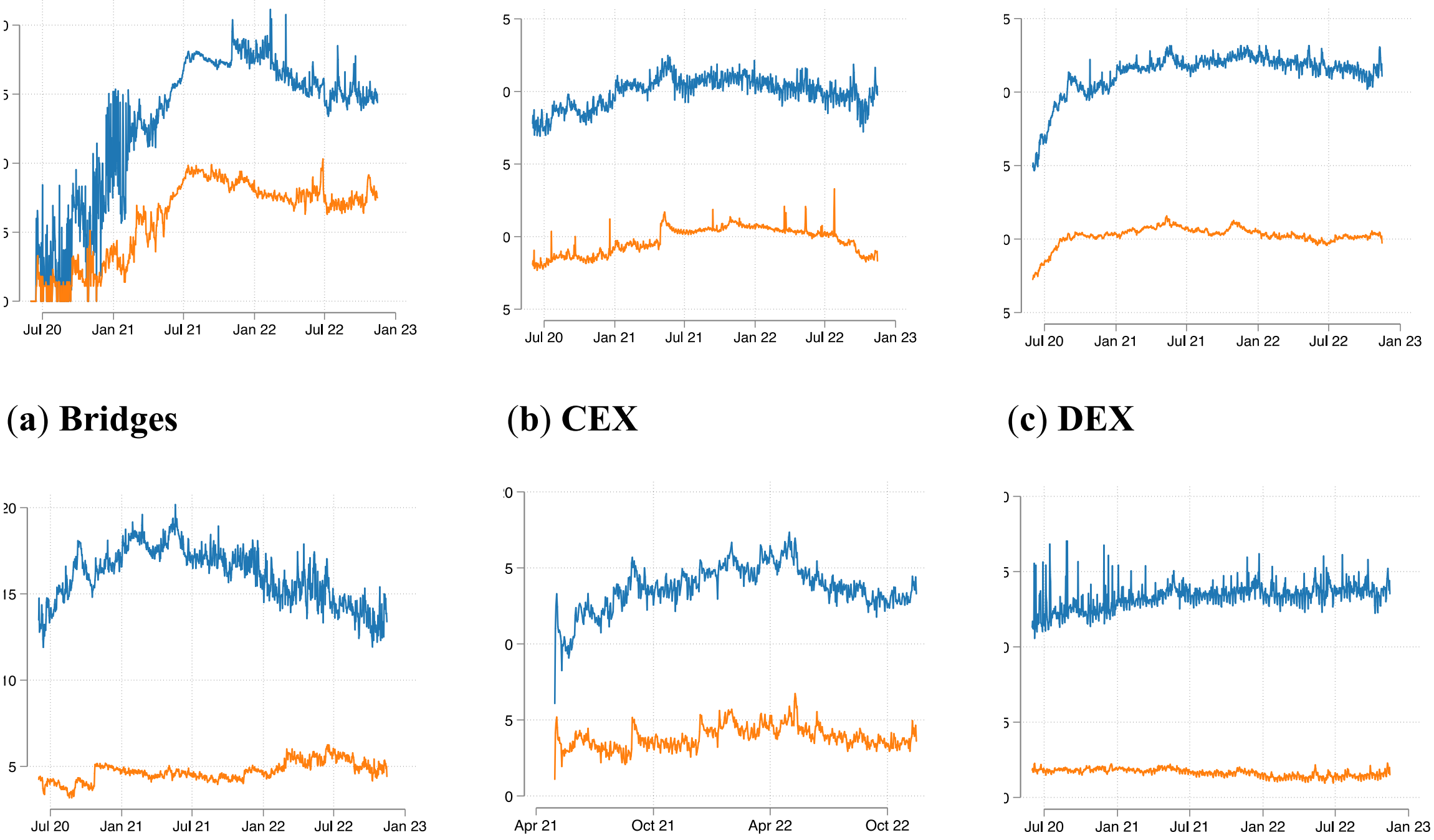

**(a) Bridges** **(b) CEX** **(c) DEX**

**(d) MEV** **(e) NFTs** **(f) Stablecoins**

***Figure 2.*** *Log-transformed transaction fees and unique active wallet users by economic system in the Ethereum ecosystem. Figure 2 features six plots that display log-transformed activity, with unique active wallets (orange) and transacted volume in USD (blue) plotted over time. The systems include (**a**) Bridges, (**b**) Centralized Exchanges (CEXs), (**c**) Decentralized Exchanges (DEXs), (**d**) Maximal Extractable Value (MEV) bots, (**e**) Non-Fungible Tokens (NFTs), and (**f**) Stablecoins. Data are obtained using Flipside Crypto. Custom application programming interfaces (APIs) were developed and implemented for each distinct contract or address on the Ethereum blockchain, allowing for the determination of unique active wallets and their corresponding transaction-associated volumes, measured in USD.*

The daily average trading volume for the analyzed (c) DEXes in Figure 2 is estimated to be approximately USD 2.9 billion. A notable expansion in volume transpires from mid-2020 to mid-2021, succeeded by a brief contraction and subsequent zenith towards the end of 2021. However, in the following year, a consistent diminution in volume materializes, interspersed with occasional fluctuations, such as a modest upswing in October 2022. This pattern may reflect evolving market conditions and user preferences. Scrutiny of the number of distinct wallets engaged in DEX transactions uncovers a pronounced increase up to mid-2021 when the apex number of unique wallets reached 105,000. Beyond this juncture, a persistent decline in user engagement manifests, punctuated by sporadic short-lived surges in activity towards the year's end. On average, 32,144 unique wallets partake in transactions with the DEXes daily throughout the examined timeframe, indicating the importance of these platforms within the decentralized finance ecosystem.

The analysis of (d) MEV volume and active wallet data in Figure 2 exhibits fluctuations, with a rapid ascent until mid-2021, succeeded by a descent. The average daily MEV volume equates to USD 26.3 million, while the mean unique active wallets number 128. A contrasting trend emerges in the unique wallet data, as a sharp upswing commences in early 2022, culminating in June 2022 at 513. This development may indicate the growing impact of MEV on blockchain security and miner incentives.

In the realm of (e) NFTs in Figure 2, the examination of the economic system commences in April 2021, corresponding with the launch of selected initiatives. The average daily trade value for NFT collections amounts to USD 2.1 million, engaging 64 distinct active wallets within the network. However, the volume exhibits substantial fluctuations (standard deviation = USD 4.1 million). The NFT volume trend reveals a steep incline until mid-2022, paralleled by a similarly sharp decline, reflecting the volatility and speculative nature of the NFT market.

The (f) stablecoins in Figure 2 underscore their significance within the Ethereum network, boasting an average daily volume of USD 22 billion and engaging over 104,000 unique active wallets. The volume remains relatively stable, punctuated by occasional upward outliers. The daily number of stablecoin users consistently surpasses 50,000 wallets, reaching its zenith of over 200,000 in November 2022, highlighting the persistent demand for stablecoins as a medium of exchange and store of value.

We employ the ADFmax test procedure with constant and trend to investigate the time series properties of the identified systems. The presence of unit roots in first differences can be rejected for all systems, as illustrated in Table 3. The unit root test suggests that the variables are optimally described in the first differences—consistent with Table 2.

**Table 3. ADFmax test results for time series properties of log-transformed economic activity by economic system in the Ethereum ecosystem.**

| | | Transaction Volume (USD) | | | Active Users | | |
|---|---|---|---|---|---|---|---|
| **System** | | **Lags** | **ADFmax** | ***p*-Value** | **Lags** | **ADFmax** | ***p*-Value** |
| (a) Bridges | AIC | 7 | −15.85 *** | 0.000 | 5 | −16.25 *** | 0.000 |
| | SIC | 5 | −20.52 *** | 0.000 | 2 | −22.99 *** | 0.000 |
| | $GTS_{05}$ | 7 | −15.85 *** | 0.000 | 5 | −16.25 *** | 0.000 |
| (b) CEX | AIC | 6 | 17.56 *** | 0.000 | 5 | −17.75 *** | 0.000 |
| | SIC | 5 | 23.80 *** | 0.000 | 5 | −17.75 *** | 0.000 |
| | $GTS_{05}$ | 5 | 23.80 *** | 0.000 | 5 | −17.75 *** | 0.000 |
| (c) DEX | AIC | 5 | −19.85 *** | 0.000 | 4 | −15.40 *** | 0.000 |
| | SIC | 5 | −19.85 *** | 0.000 | 1 | −25.49 *** | 0.000 |
| | $GTS_{05}$ | 5 | −19.85 *** | 0.000 | 4 | −15.40 *** | 0.000 |
| (d) MEV | AIC | 6 | −15.84 *** | 0.000 | 2 | −22.57 *** | 0.000 |
| | SIC | 4 | −19.05 *** | 0.000 | 2 | −22.57 *** | 0.000 |
| | $GTS_{05}$ | 6 | −15.84 *** | 0.000 | 2 | −22.57 *** | 0.000 |
| (e) NFTs | AIC | 3 | −17.27 *** | 0.000 | 3 | −19.39 *** | 0.000 |
| | SIC | 0 | −33.10 *** | 0.000 | 3 | −19.39 *** | 0.000 |
| | $GTS_{05}$ | 3 | −17.27 *** | 0.000 | 3 | −13.39 *** | 0.000 |
| (f) Stablecoins | AIC | 7 | −17.92 *** | 0.000 | 7 | −12.02 *** | 0.000 |
| | SIC | 6 | −20.63 *** | 0.000 | 6 | −14.03 *** | 0.000 |
| | $GTS_{05}$ | 6 | −20.63 *** | 0.000 | 6 | −14.03 *** | 0.000 |

*Table 3 presents the results of an ADFmax test, incorporating a constant and trend for both log-transformed and first-differenced unique active wallets and transaction volumes in USD across various economic subsystems within the Ethereum blockchain. The systems include (a) Bridges, (b) Centralized Exchanges (CEXs), (c) Decentralized Exchanges (DEXs), (d) Maximal Extractable Value (MEV) bots, (e) Non-Fungible Tokens (NFTs), and (f) Stablecoins. Significance levels are denoted by *, **, and *** for 10%, 5%, and 1%, respectively.*

## *3.3. Empirical Approach*

### 3.3.1. Granger Causality

Granger causality is a statistical concept that aims to determine the causal relationship between two time series variables. The concept of Granger causality states that if the past values of variable $y_1$ can be used to predict the current value of variable $y_2$, taking into account the past values of $y_2$, then $y_1$ is said to Granger cause $y_2$ (Granger 1981; Engle and Granger 1987). The formal approach described in the following is adapted from the method described by Baum et al. (2021, 2022) to analyze the temporal stability of Granger-causal relationships. In a bivariate VAR(m) model, $y_{1t}$ and $y_{2t}$ represent economic time series of interest.

$$y_{1t} = \emptyset_0^{(1)} \sum_{k=1}^{m} \emptyset_{1k}^{(1)} y_{1t-k} + \sum_{k=1}^{m} \emptyset_{2k}^{(1)} y_{2t-k} + \varepsilon_{1t}, \text{ and} \quad (1)$$

$$y_{2t} = \emptyset_0^{(2)} \sum_{k=1}^{m} \emptyset_{1k}^{(2)} y_{1t-k} + \sum_{k=1}^{m} \emptyset_{2k}^{(2)} y_{2t-k} + \varepsilon_{2t}, \quad (2)$$

The joint significance of multiple parameters is evaluated using a Wald test, examining the null hypothesis of no causality from $y_1$ to $y_2$. The system can be reshaped in matrix notation, with $y_t$ = [$y_{1t}$ $y_{2t}$]', $x_t$ = [1 $y'_{t-1}$ $y'_{t-2}$ … $y'_{t-k}$]', and ∏ 2$x$ (2$m$ + 1) = [Φ0 Φ1 … Φm] with Φ0 = [$\emptyset_0^{(1)}$ $\emptyset_0^{(2)}$]' and $\Phi_k = \begin{bmatrix} \emptyset_{1k}^{(1)} & \emptyset_{2k}^{(1)} \\ \emptyset_{1k}^{(2)} & \emptyset_{2k}^{(2)} \end{bmatrix}$ for $k$ = 1, …, $m$.

Thus, the bivariate VAR(m) can be expressed as:

$$y_t = \prod x_t + \varepsilon_t. \quad (3)$$

The null hypothesis of the absence of causality from $y_1$ to $y_2$ is represented by $R_{1\to 2}\pi = 0$, where $R_{1\to 2}$ serves as the coefficient restriction matrix and π = vec(∏). The statistic used to evaluate this null hypothesis in the presence of heteroskedasticity is referred to as $W_{1\to 2}$ and is calculated as:

$$W_{1\to 2} = T(R_{1\to 2}\hat{\pi})', \left[R_{1\to 2}\left(\hat{V}^{-1}\hat{\Sigma}\hat{V}^{-1}\right)R'_{1\to 2}\right]^{-1}(R_{1\to 2}\hat{\pi}), \quad (4)$$

with $\hat{V} = I_N \otimes \hat{Q}$ and $\hat{Q} = T^{-1}\sum_t x_t x'_t$ and $\hat{\Sigma} = T^{-1}\sum_t \xi_t \xi'_t$. The term $\xi_t$ stands for $\hat{\varepsilon} \otimes x_t$, with $\hat{\varepsilon}_t = y_t - \hat{\prod} x_t$. $I_N$ refers to the number of variables in the VAR model.

The framework for testing Granger causality within the context of a VAR model, estimated using stationary variables, is augmented to account for the possibility of integrated variables. Toda and Yamamoto (1995) and Dolado and Lütkepohl (1996) suggest the use of a Lag-Augmented VAR (LA-VAR) model, which is an extension of the original VAR(m) model with the inclusion of d lags to account for the maximum order of integration of the variables. The resulting model is represented as VAR($m$ + $d$).

The procedure for conducting a Granger causality test within the framework of an LA-VAR model remains unchanged, with the exception that the coefficients associated with the additional d lags are not included in the testing restrictions.

### 3.3.2. Time-Varying Granger Causality

The validity and robustness of VAR results are often contingent upon the specific time period over which the VAR is estimated. This highlights the need for a more comprehensive approach when assessing structural stability, as the existence of Granger causality between a pair of variables may be supported over one time frame, yet may be found to be fragile when alternative periods are considered. Recent literature, such as the work of Phillips et al. (2011, 2014, 2015a, 2015b) has contributed significantly to the field by developing methods for detecting and dating financial bubbles. These methods involve the use of right-tailed unit root tests in conjunction with date-stamping techniques. Subsequently, the concept of Granger causality has been extended by Shi et al. (2018, 2020) to incorporate these techniques, resulting in a more robust approach to assessing causality.

In order to analyze time-varying Granger causality, recursive estimation methods are required. These methods involve computing a sequence of test statistics for each time period of interest and using this information for inference. The three algorithms that generate this sequence of test statistics are the forward expanding window (FE) (Thoma 1994), the rolling window (RO) (Arora and Shi 2016; Swanson 1998), and the recursive evolving (RE) algorithms (Shi et al. 2020; Phillips et al. 2015a). These algorithms are illustrated in Figure 3, where each arrow represents a subsample over which the test statistic is computed. Given a sample of $T + 1$ observations, denoted as $\{y_0, y_1, \ldots, y_T\}$, and a value r such that $0 < r < 1$, the Wald test statistic is computed over a subsample starting at $y[Tr_1]$ and ending at $y[Tr]$.

The interpretation of causality (or lack thereof) in the results derived from various algorithms is predicated on the premise that a minimum of two out of three tests must exhibit concordant outcomes.

This criterion serves as an indicator of causal influences. The application of three distinct tests provides a validation of the robustness of the results.

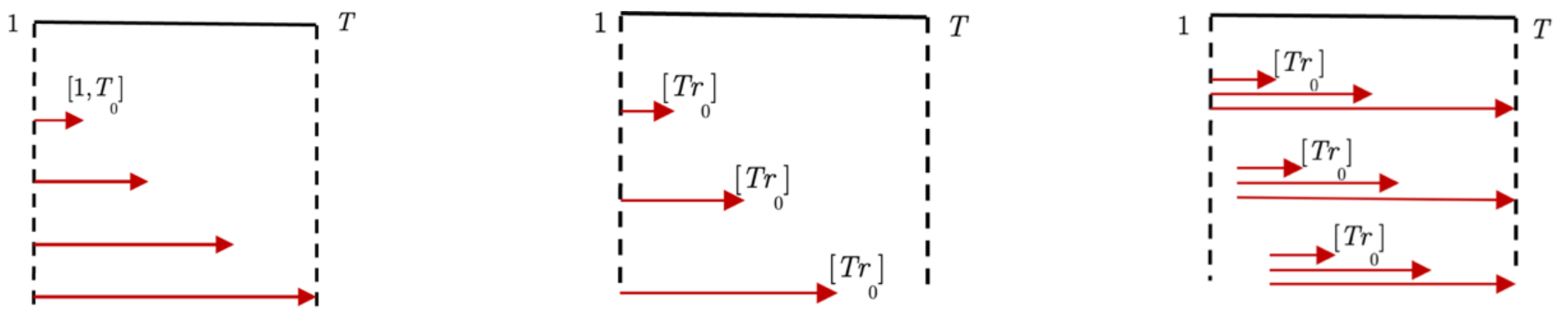

**(a) Forward expanding window** **(b) Rolling window** **(c) Recursive evolving window**

***Figure 3.*** *Sample sequences and window widths. Adapted from Phillips et al. (2015a). The red arrows indicate the temporal scope and progression for each test statistic. For the Forward Expanding Window (FE), the arrow originates from the initial data point and moves forward, reflecting an increasing window as new data is included. The Rolling Window (RO) displays arrows of equal length, symbolizing a constant-sized window that progresses through the data, discarding the oldest point for a new one at each step. Lastly, the Recursive Evolving Window (RE) shows arrows beginning at successive data points, each extending to the end, depicting windows that expand over time starting from various points in the series.*

# 4. Results

## *4.1. Baseline Estimation*

Table 4 presents the Wald statistic outcomes and the 95% and 99% thresholds for inferring causality among economic systems within the Ethereum ecosystem, including (a) bridge volume and bridge activity; (b) CEX volume and CEX activity; (c) DEX volume and DEX activity; (d) MEV volume and MEV activity; (e) NFT volume and NFT activity; and (f) stablecoin volume and stablecoin activity.

**Table 4. Time-varying Granger causality estimates between economic systems on the Ethereum blockchain.**

| | Forward | | | Rolling | | | Recursive | | |
|---|---|---|---|---|---|---|---|---|---|
| **Direction of Causality** | **Wald** | **95th** | **99th** | **Wald** | **95th** | **99th** | **Wald** | **95th** | **99th** |
| (a) Bridges | | | | | | | | | |
| Bridge volume $\overset{GC}{\to}$ Fees | 7.09 | 8.98 | 14.18 | 16.73 *** | 9.86 | 15.08 | 16.81 *** | 10.19 | 15.56 |
| Bridge activity $\overset{GC}{\to}$ Fees | 17.20 *** | 9.48 | 15.63 | 19.71 *** | 8.71 | 15.41 | 27.63 *** | 11.03 | 16.04 |
| Fees $\overset{GC}{\to}$ Bridge volume | 4.13 | 7.19 | 11.29 | 15.47 *** | 7.68 | 12.87 | 15.67 *** | 7.87 | 12.87 |
| Fees $\overset{GC}{\to}$ Bridge activity | 9.19 ** | 6.22 | 11.92 | 13.95 *** | 6.45 | 11.84 | 15.89 *** | 6.82 | 11.92 |
| (b) CEX | | | | | | | | | |
| CEX volume $\overset{GC}{\to}$ Fees | 27.05 *** | 8.17 | 11.47 | 31.87 *** | 9.63 | 15.32 | 47.96 *** | 10.27 | 15.63 |
| CEX activity $\overset{GC}{\to}$ Fees | 3.42 | 7.39 | 13.27 | 46.99 *** | 8.74 | 15.85 | 46.99 *** | 9.06 | 16.04 |
| Fees $\overset{GC}{\to}$ CEX volume | 52.35 *** | 7.69 | 14.42 | 29.24 *** | 9.03 | 13.82 | 55.79 *** | 9.29 | 14.51 |
| Fees $\overset{GC}{\to}$ CEX activity | 16.65 *** | 7.46 | 11.31 | 33.49 *** | 7.65 | 13.53 | 33.49 *** | 7.94 | 14.51 |
| (c) DEX | | | | | | | | | |
| DEX volume $\overset{GC}{\to}$ Fees | 9.07 ** | 8.93 | 15.08 | 32.20 *** | 10.62 | 14.52 | 32.39 *** | 10.97 | 15.46 |
| DEX activity $\overset{GC}{\to}$ Fees | 30.77 *** | 9.07 | 14.38 | 29.33 *** | 9.94 | 15.47 | 36.66 *** | 10.60 | 15.86 |
| Fees $\overset{GC}{\to}$ DEX volume | 17.36 *** | 9.17 | 15.60 | 19.52 *** | 9.39 | 17.92 | 36.40 *** | 9.70 | 17.93 |
| Fees $\overset{GC}{\to}$ DEX activity | 18.98 *** | 10.85 | 16.11 | 20.69 *** | 11.97 | 15.36 | 22.04 *** | 11.98 | 16.17 |
| (d) MEV | | | | | | | | | |
| MEV volume $\overset{GC}{\to}$ Fees | 13.85 ** | 11.03 | 23.09 | 18.27 ** | 15.86 | 25.40 | 18.31 ** | 16.65 | 25.99 |
| MEV activity $\overset{GC}{\to}$ Fees | 13.41 | 13.59 | 23.09 | 21.54 *** | 13.83 | 24.30 | 26.06 *** | 14.58 | 26.01 |
| Fees $\overset{GC}{\to}$ MEV volume | 13.21 ** | 11.36 | 17.52 | 20.01 *** | 12.06 | 17.01 | 23.43 *** | 12.51 | 17.52 |
| Fees $\overset{GC}{\to}$ MEV activity | 8.56 | 11.20 | 19.01 | 21.51 *** | 11.49 | 18.74 | 21.62 *** | 12.26 | 19.01 |
| (e) NFT | | | | | | | | | |
| NFT volume $\overset{GC}{\to}$ Fees | 6.94 | 8.19 | 14.33 | 17.42 ** | 9.03 | 18.90 | 17.42 ** | 9.61 | 18.90 |
| NFT activity $\overset{GC}{\to}$ Fees | 11.58 ** | 9.00 | 12.50 | 17.67 *** | 9.03 | 12.50 | 23.13 *** | 9.71 | 13.40 |
| Fees $\overset{GC}{\to}$ NFT volume | 12.57 ** | 10.25 | 13.79 | 37.30 *** | 10.64 | 13.90 | 41.21 *** | 10.67 | 14.15 |
| Fees $\overset{GC}{\to}$ NFT activity | 12.39 | 12.82 | 16.16 | 36.77 *** | 13.02 | 17.99 | 38.66 *** | 13.46 | 17.99 |
| (f) Stablecoins | | | | | | | | | |
| Stablecoin volume $\overset{GC}{\to}$ Fees | 9.36 *** | 7.05 | 9.31 | 11.52 ** | 8.03 | 11.79 | 16.79 *** | 8.07 | 12.92 |
| Stablecoin activity $\overset{GC}{\to}$ Fees | 13.12 ** | 12.49 | 16.90 | 26.84 *** | 13.83 | 20.67 | 40.98 *** | 14.50 | 20.67 |
| Fees $\overset{GC}{\to}$ Stablecoin volume | 9.07 ** | 8.58 | 11.28 | 23.61 *** | 8.55 | 11.73 | 36.78 *** | 8.93 | 11.90 |
| Fees $\overset{GC}{\to}$ Stablecoin activity | 41.55 *** | 7.43 | 12.84 | 18.11 *** | 9.86 | 13.66 | 43.56 *** | 9.14 | 13.66 |

*Table 4 presents the robust Wald test statistics for Granger causality tests, with the 95th and 99th quantiles of the empirical distributions based on 499 bootstrap replications. *, **, and *** for 10%, 5%, and 1%, respectively.*

The empirical analysis uncovers statistically significant bivariate causal interconnections amongst all variables, employing rolling (RO) and recursive evolving (RE) algorithms. This highlights the crucial role of various forms of economic activities in shaping the Ethereum network's dynamics. In certain instances, the forward expanding (FE) algorithm yields statistically insignificant results; however, Phillips et al. (2015a) posit that the forward algorithm exhibits less reliability compared to its rolling and recursive counterparts.

### *4.2. Bridges*

The interplay between bridge utilization and transaction fees constitutes a subject of considerable interest within the domain of blockchain economics. Blockchain bridges, as technological solutions, enable interoperability among disparate blockchain networks. The connection between these bridges and the Ethereum network may be subject to various influences, such as a potential decline in Ethereum network activity attributable to high bridge usage, or a fee increase driven by elevated demand for bridge services. It is reasonable to anticipate that blockchain bridges influence Ethereum transaction fees by alleviating network congestion, fostering competition among blockchain networks, facilitating transaction fee arbitrage opportunities, and promoting layer-2 scaling solutions. Intuitively, higher fees may prompt certain users to explore alternative, lower-fee blockchain network options, while a fee reduction could, in theory, entice certain users to rejoin the Ethereum network. Nevertheless, the migration process could culminate in the permanent departure of projects or users, as they may be reluctant to face the risk of potential future fee hikes and the associated need for subsequent migration, or they might simply be content with another blockchain network. The presence of bridging services and solutions, such as Axie Infinity's Ronin Bridge, exemplifies the correlation between transaction costs and migration within the blockchain ecosystem. The Ronin Bridge, for instance, is a smart contract that allows users of the NFT or play-to-earn game Axie Infinity to transition to the Ronin sidechain, which was primarily established due to the elevated transaction costs on the Ethereum network (Axie 2021).

The results of the time-varying Granger causality analysis between bridge transaction volume, activity, and mean fees in the Ethereum network, are illustrated in Figure 4. The dashed lines denote the critical values of bootstrapped test statistics at the 90% and 95% levels. A Granger curve positioned above these lines indicates the presence of significant causal relationships in the context of Ethereum transaction fees. Figure 4 reveals that Ethereum fees generally exert a Granger-causal influence on bridge activity over time (Figure 4, panels d,h,l). This suggests that fluctuations in fees directly impact user behavior on bridges. Notably, a feedback loop is observed, as bridge activity also exhibits a Granger-causal effect on Ethereum fees until Q2 of 2022 (Figure 4, panels b,f,j). This observation implies that users may have migrated to alternative blockchains, consequently weakening the causal relationship between these variables—consistent with our expectations. Despite a notable decrease in average fees over the sample period (see Figure 1), the causal influence of fees on bridge activity seems to persist throughout much of the sample (Figure 4, panels b,j). This observation suggests that the fees in the Ethereum network, despite being comparatively low, may not be competitive enough in comparison to other networks and layer-2 solutions, leading to a continued churn in the network. In contrast, the analysis generally does not reveal significant causal effects of Ethereum fees on bridge transaction volume (Figure 4, panels c,g,k). The findings concerning the causal impact of bridge transaction volume on Ethereum fees also yield mixed results (Figure 4, panels a,e,i).

### *4.3. Centralized Exchanges (CEX)*

The relationship between CEX hot wallet movements and Ethereum transaction fees can be delineated through an examination of several distinct factors such as network congestion, arbitrage opportunities, and exchange operations. Network congestion arises when CEXs process numerous transactions, increasing demand for block space on the Ethereum blockchain and intensifying competition for transaction processing. In addition, arbitrage opportunities are created due to price discrepancies between CEXs and other exchanges or DEXs, prompting traders to buy low on one exchange and sell high on another. These activities require asset transfers between wallets, contributing to hot wallet movements and higher transaction fees. Additionally, CEXs routinely transfer funds between their hot and cold wallets for security and operational purposes, further impacting Ethereum transaction fees.

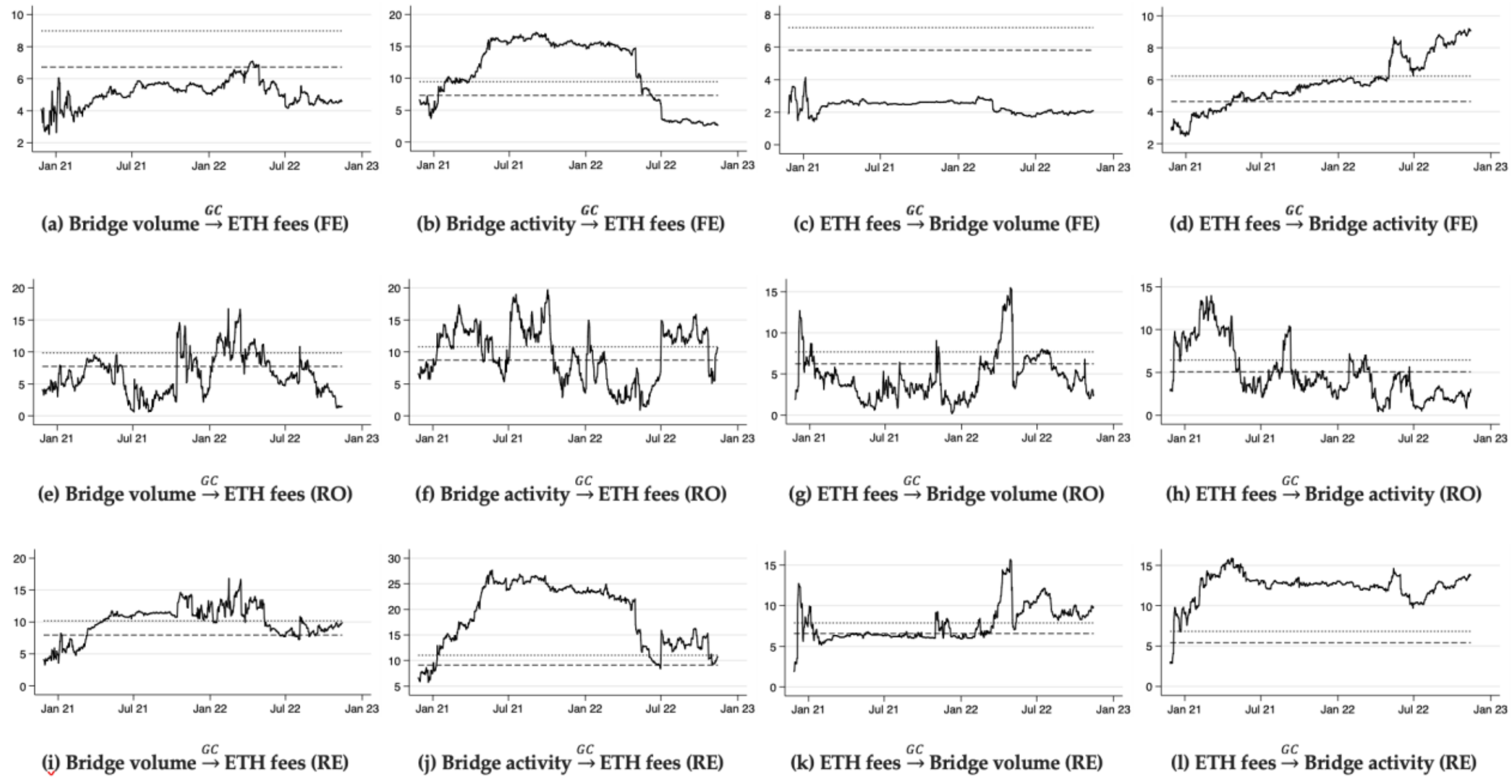

***Figure 4.*** *Time-varying Granger causality tests for bridge volume, activity, and mean fees in the Ethereum network. Figure 4 displays the bivariate results derived from forward expanding (FE), rolling (RO), and recursive evolving (RE) algorithms using the time-varying Granger causality model developed by Baum et al. (2021, 2022). The analysis employs the Lag-Augmented Vector Autoregression (LA-VAR) model proposed by Toda and Yamamoto (1995) and Dolado and Lütkepohl (1996). The sample period spans from 1 July 2020 to 14 November 2022, with a minimum window size set at 20% of the sample. Models incorporate four augmented lags and a trend. Dashed lines represent the critical values of bootstrapped test statistics at the 90% and 95% significance levels. The results are robust to heteroskedasticity.*

The time-varying Granger causality analysis between CEX transaction volume, activity, and mean fees in the Ethereum network, as depicted in Figure 5, reveals a mutual interaction between the variables. Ethereum fees generally exert a Granger-causal influence on CEX transaction volume, with the causal effect growing in statistical significance over time (Figure 5, panels c,g,k). A feedback loop is also observed, as CEX transaction volumes typically Granger-cause Ethereum fees (Figure 5, panels a,e,i). This suggests that as CEX transaction volume increases, it spurs heightened demand for block space and greater competition among users for transaction processing. Consequently, users are willing to pay higher gas fees to expedite their transactions, driving up average Ethereum transaction fees. When Ethereum transaction fees increase, they affect trading and transferring costs on CEXs, influencing user behavior. This may incentivize traders to engage in more high-value transactions, leading to a rise in CEX transaction volume and reinforcing the causal relationship between Ethereum fees and CEX transaction volume. The presence of a feedback loop implies that market participants closely monitor the interplay between Ethereum transaction fees and CEX transaction volume, transacting more on CEXs in response to increasing Ethereum fees to capitalize on arbitrage opportunities or market volatility. This, in turn, contributes to network congestion and further elevates Ethereum transaction fees in a self-reinforcing cycle. Overall, the analysis uncovers a bidirectional relationship between CEX transaction volume and Ethereum fees that strengthens over time, driven primarily by supply and demand dynamics for block space on the Ethereum network and the strategic behavior of market participants in response to fluctuating transaction costs.

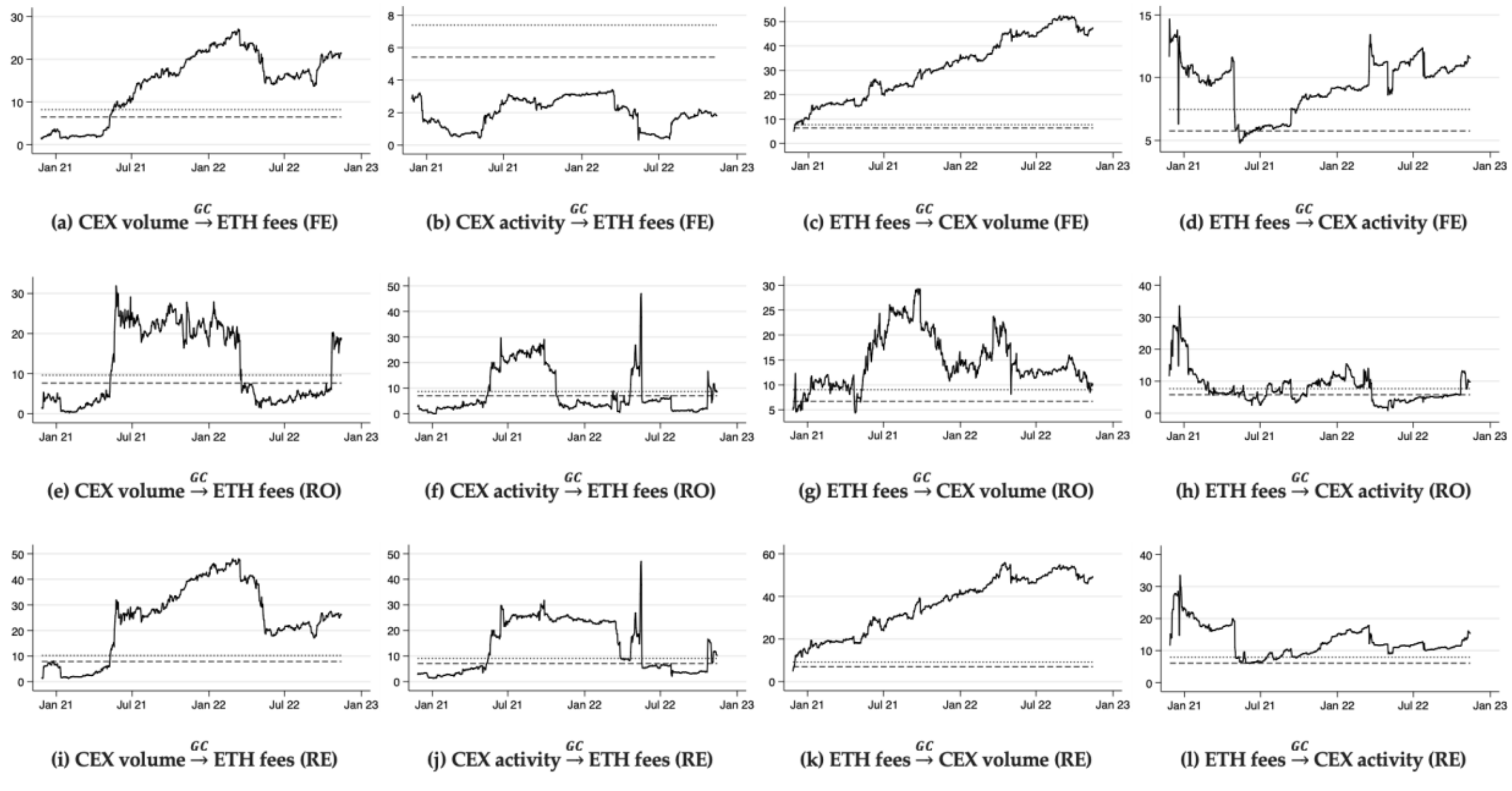

***Figure 5.*** *Time-varying Granger causality tests for CEX volume, activity, and mean fees in the Ethereum network. Figure 5 displays the bivariate results derived from forward expanding (FE), rolling (RO), and recursive evolving (RE) algorithms using the time-varying Granger causality model developed by Baum et al. (2021, 2022). The analysis employs the Lag-Augmented Vector Autoregression (LA-VAR) model proposed by Toda and Yamamoto (1995) and Dolado and Lütkepohl (1996). The sample period spans from 1 July 2020 to 14 November 2022, with a minimum window size set at 20% of the sample. Models incorporate four augmented lags and a trend. Dashed lines represent the critical values of*

*bootstrapped test statistics at the 90% and 95% significance levels. The results are robust to heteroskedasticity.*

### *4.4. Decentralized Exchanges (DEX)*

DEXs enable peer-to-peer trading of digital assets without a centralized intermediary by leveraging smart contracts and automated market-making protocols, fostering direct, trustless transactions between participants. User activity on DEXs is influenced by factors such as market sentiment, profitable trading opportunities, and the expansion of the decentralized finance (DeFi) ecosystem. As user activity and transaction volumes increase, the demand for limited resources on the underlying blockchain network rises, intensifying competition for block space and leading to higher transaction fees. In this context, elevated transaction fees on networks like Ethereum may deter DEX users from conducting trades or interacting with decentralized applications (dApps), causing a decline in transaction volumes and user activity until fees revert to lower levels. Conversely, reduced transaction fees on Ethereum may incentivize users to engage in trades and other activities on DEXs, thereby boosting transaction volumes and user activity. Ultimately, the interplay between transaction volumes, user activity on DEXs, and Ethereum transaction fees reflects a dynamic relationship, with increased demand for network resources contributing to fluctuating transaction fees.

The time-varying Granger causality analysis, presented in Figure 6, explores the relationship between DEX transaction volumes, activity, and mean fees in the Ethereum network. The results reveal that, from Q3 2022 onwards, DEX transaction volumes generally Granger-caused Ethereum transaction fees (Figure 6, panels e,i). Over the same period, Ethereum transaction fees did not consistently exhibit a Granger-causal effect on DEX transaction volumes (Figure 6, panels c,g,k), although sporadic instances of Granger causality between Ethereum transaction fees and DEX activity are observed (Figure 6, panels d,h,l). The Granger-causal relationship between DEX volume and Ethereum transaction fees can be attributed to the increasing utilization of DEXs, which amplifies demand for limited resources on the Ethereum blockchain network. Consequently, escalating transaction fees are driven by heightened competition for block space due to increased user activity and transaction volumes on DEXs.

However, the lack of a consistent Granger-causal relationship between Ethereum transaction fees and DEX volumes suggests that multiple factors, such as market sentiment, trading opportunities, and the expansion of the DeFi ecosystem, influence users' behavior beyond transaction fees alone. Therefore, the impact of transaction fees on DEX volumes is more nuanced, with external factors potentially playing a more significant role in determining DEX activity. The occasional instances where Ethereum transaction fees Granger-cause DEX activity emphasize the complex and dynamic nature of the relationship between Ethereum transaction fees and DEX activity, highlighting periods when elevated transaction fees might have a more pronounced influence on user behavior, discouraging users from engaging with DEXs and causing a decline in transaction volumes and user activity.

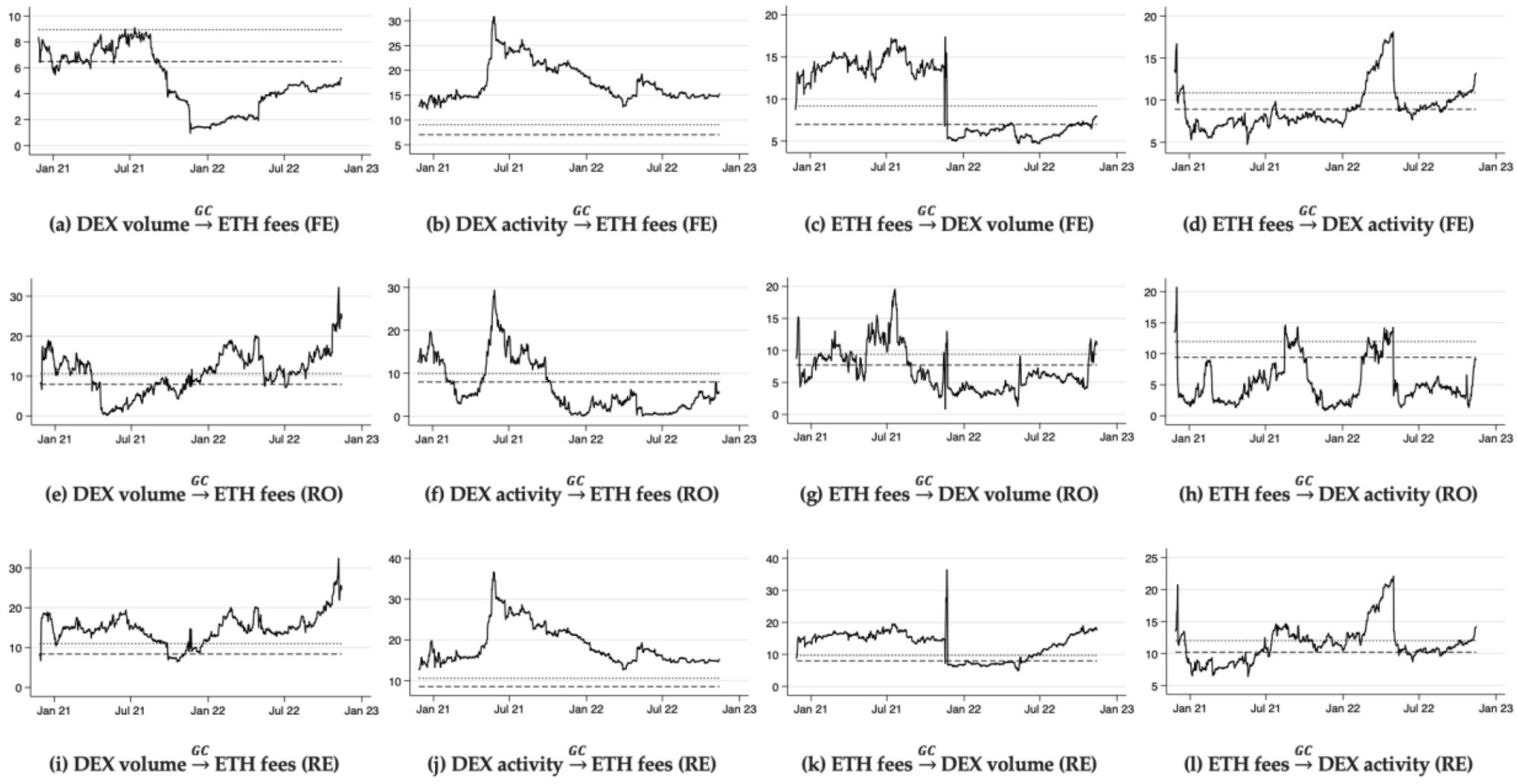

***Figure 6.*** *Time-varying Granger causality tests for DEX volume, activity, and mean fees in the Ethereum network. Figure 6 displays the bivariate results derived from forward expanding (FE), rolling (RO), and recursive evolving (RE) algorithms using the time-varying Granger causality model developed by Baum et al. (2021, 2022). The analysis employs the Lag-Augmented Vector Autoregression (LA-VAR) model proposed by Toda and Yamamoto (1995) and Dolado and Lütkepohl (1996). The sample period spans from 1 July 2020 to 14 November 2022, with a minimum window size set at 20% of the sample. Models incorporate four augmented lags and a trend. Dashed lines represent the critical values of bootstrapped test statistics at the 90% and 95% significance levels. The results are robust to heteroskedasticity.*

### *4.5. Maximal Extractable Value (MEV)*

Prior to the Merge, Maximal Extractable Value (MEV) referred to the additional value that could be obtained by miners in a proof-of-work (PoW) consensus mechanism by strategically including, excluding, or ordering transactions in a block. After the Merge, when Ethereum transitioned to a proof-of-stake (PoS) consensus mechanism, the term Maximal Extractable Value replaced Miner Extractable Value. In this new context, validators took on the role of managing transaction inclusion, exclusion, and ordering, as opposed to miners. Although the terminology and consensus mechanism changed, the core concept of extracting additional value from block production beyond standard block rewards and transaction fees remained consistent.

The interrelation between MEV activity, characterized by the number of searchers pursuing profitable opportunities, and Ethereum transaction fees is driven by the competitive nature of MEV extraction. Heightened MEV activity intensifies competition for transaction inclusion in blocks, prompting searchers to offer higher gas fees to validators to increase the likelihood of their transactions being included and, consequently, raising the MEV rewards. This competitive bidding process results in elevated Ethereum transaction fees. Moreover, MEV transaction volumes, representing the cumulative value derived from MEV opportunities, play a significant role in determining Ethereum transaction fees. As MEV transaction volumes grow, searchers are drawn to the potential rewards, exacerbating competition and driving gas prices upward. In highly competitive MEV scenarios, such as DEX arbitrage, searchers may allocate up to 90% or more of their total MEV revenue to gas fees to ensure transaction inclusion in a block (Ethereum Foundation 2023). Furthermore, the deployment of generalized frontrunner bots by some searchers adds to the competitive environment for transaction inclusion. These bots monitor the Mempool for lucrative transactions, replicate transaction codes, replace addresses with their own, and resubmit the modified transaction with higher gas prices. This frontrunning phenomenon contributes to increased Ethereum transaction fees by further intensifying the competition for transaction inclusion.

The time-varying Granger causality analysis, presented in Figure 7, explores the relationship between MEV transaction volumes, activity, and mean fees in the Ethereum network. The results reveal that, from Q1 2022, Ethereum transaction fees generally Granger-caused MEV activity (Figure 7, panels h,l). However, from Q2 2022 onwards, the Granger-causal influence of Ethereum transaction fees on MEV activity (Figure 7, panels c,g,k) weakened in significance. Likewise, the impact of MEV activity (Figure 7, panels a,e,i) and volume (Figure 7, panels b,f,j) on Ethereum transaction fees also declined in significance from Q2 2022 onwards. This reduction in significance could be attributed to several factors, such as the emergence of alternative value extraction methods, changes in searcher strategies, or adaptations in the Ethereum ecosystem. Additionally, market participants might have adjusted their behaviors in response to the evolving dynamics of MEV opportunities and transaction fees, leading to new equilibrium points in the competitive landscape.

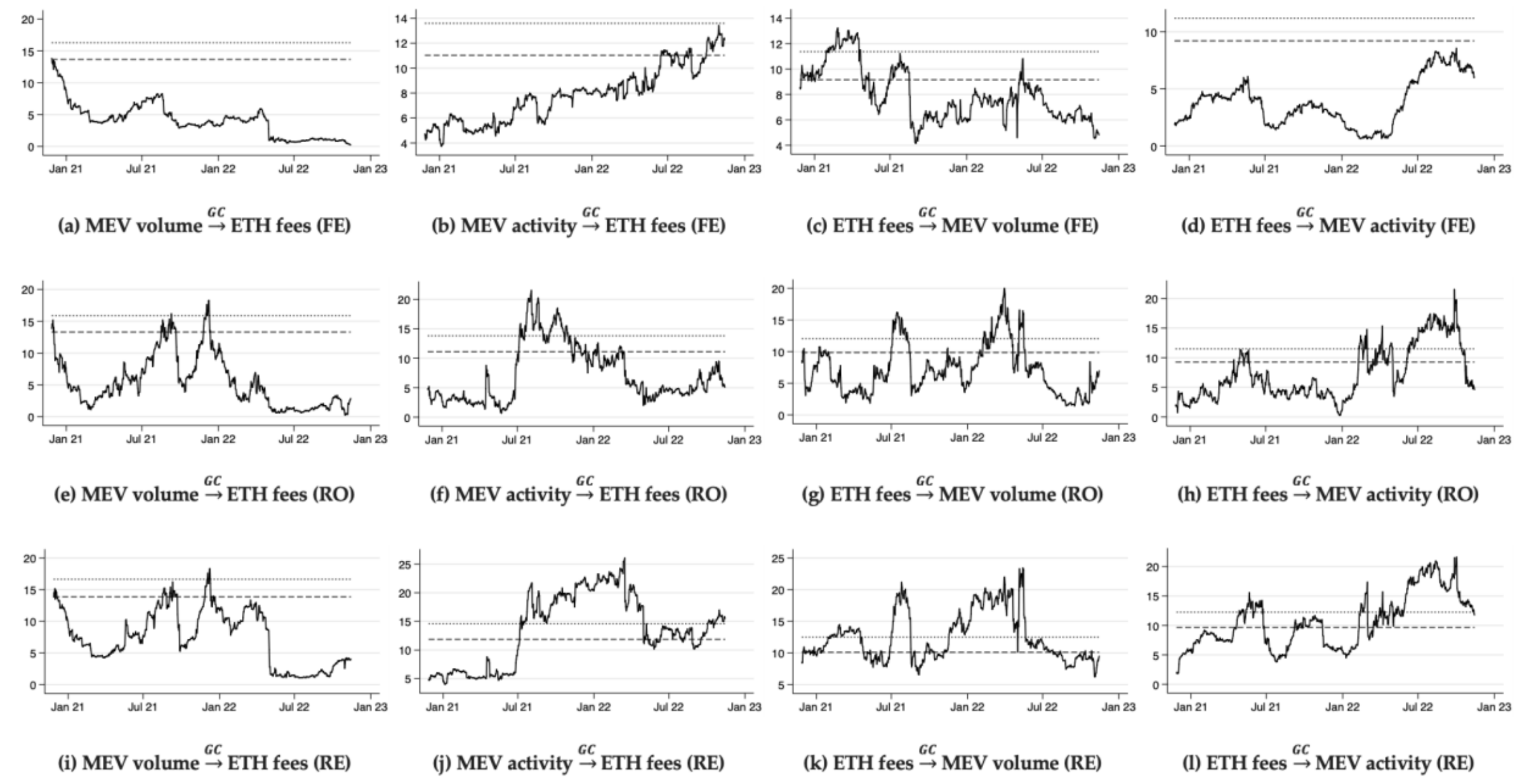

***Figure 7.*** *Time-varying Granger causality tests for MEV volume, activity, and mean fees in the Ethereum network. Figure 7 displays the bivariate results derived from forward expanding (FE), rolling (RO), and recursive evolving (RE) algorithms using the time-varying Granger causality model developed by Baum et al. (2021, 2022). The analysis employs the Lag-Augmented Vector Autoregression (LA-VAR) model proposed by Toda and Yamamoto (1995) and Dolado and Lütkepohl (1996). The sample period spans from 1 July 2020 to 14 November 2022, with a minimum window size set at 20% of the sample. Models incorporate four augmented lags and a trend. Dashed lines represent the critical values of bootstrapped test statistics at the 90% and 95% significance levels. The results are robust to heteroskedasticity.*

### *4.6. Non-Fungible Tokens (NFTs)*

NFTs represent a novel class of digital assets, characterized by their unique and non-interchangeable properties, which distinguish them from fungible tokens such as cryptocurrency coins and tokens. Primarily developed on the Ethereum blockchain utilizing ERC-721 or ERC-1155 token standards, NFTs enable the tokenization of a diverse range of digital and physical assets, including digital art, collectibles, virtual real estate, and tangible goods. As NFT transactions contribute to the overall transaction volume on the Ethereum network, escalating NFT transaction volumes and non-fungible user activity, which encompass actions like minting, trading, and transferring NFTs, can intensify network congestion. During periods of high demand, competition for limited block space heightens, leading users to offer higher gas prices and subsequently drive up Ethereum transaction fees. The notable influence of NFT volume on fees in April and September 2022 can be ascribed to speculative bubbles prevalent in the NFT market, as corroborated by the literature (Maouchi et al. 2022; Wang et al. 2022).

The time-varying Granger causality analysis between NFT transaction volume, activity, and mean fees in the Ethereum network is presented in Figure 8. Estimates reveal that Ethereum transaction fees primarily Granger-caused NFT activity (Figure 8, panels h,l) and transaction volumes (Figure 8, panels g,k) between Q4 2021 and Q1 2022. One possible interpretation of this causal relationship is that higher transaction fees might have served as a barrier to entry for some users, discouraging them from engaging in NFT-related activities and leading to reduced transaction volumes. Conversely, lower transaction fees may have acted as a catalyst for NFT user activity, incentivizing users to participate in minting, trading, and transferring NFTs, thereby increasing transaction volumes. However, the evidence for NFT user activity (Figure 8, panels f,j) or transaction volume (Figure 8, panels a,e,i) Granger-causing Ethereum transaction fees are mixed, suggesting a more nuanced causal relationship. This absence of a consistent causal link may be attributed to various confounding factors impacting Ethereum transaction fees, such as network capacity, miner preferences, and the overall transaction demand on the Ethereum network. For instance, a surge in non-NFT-related transactions may have led to increased network congestion, subsequently driving up transaction fees, independent of NFT user activity or transaction volume.

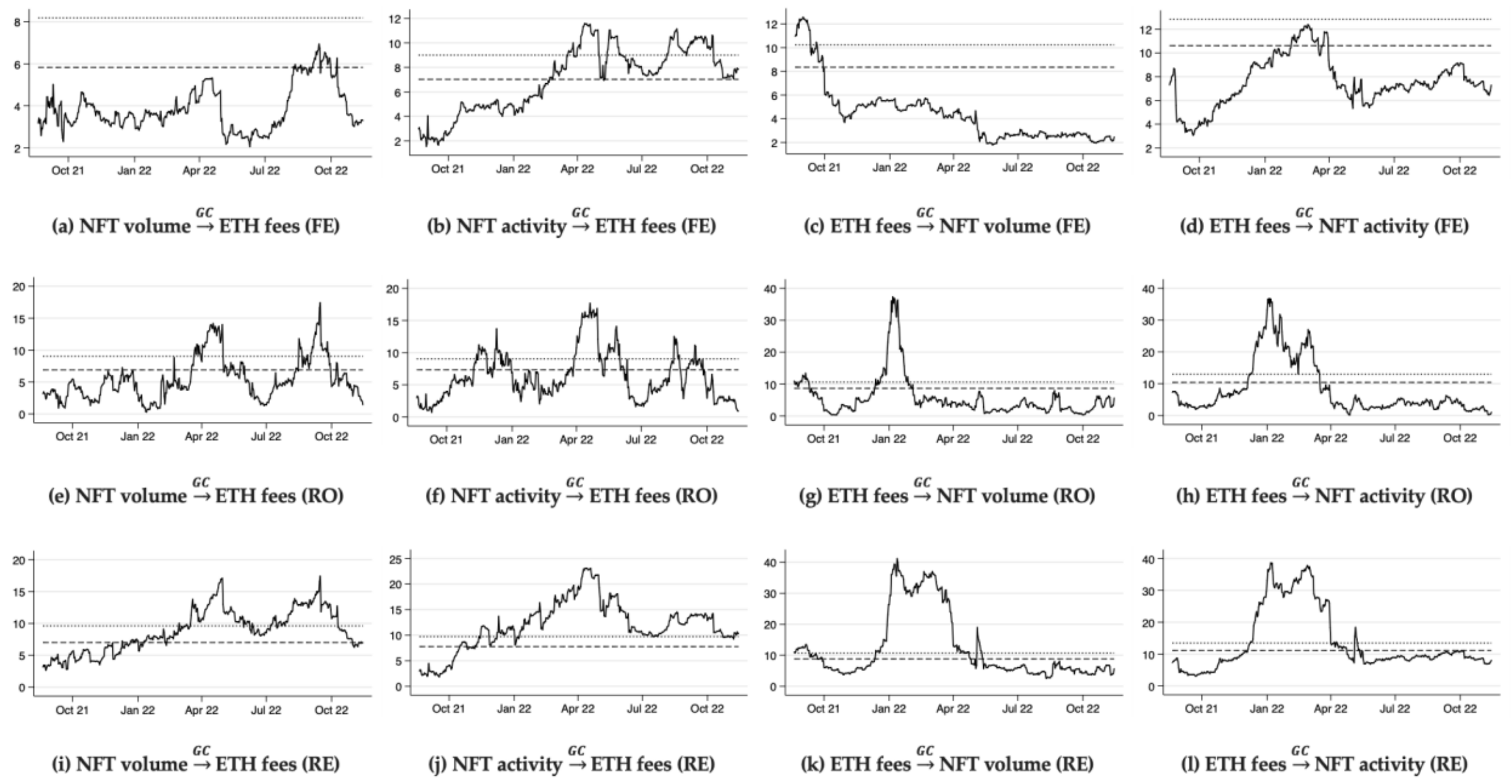

***Figure 8.*** *Time-varying Granger causality tests for NFT volume, activity, and mean fees in the Ethereum network. Notes: Figure 8 displays the bivariate results derived from forward expanding (FE), rolling (RO), and recursive evolving (RE) algorithms using the time-varying Granger causality model developed by Baum et al. (2021, 2022). The analysis employs the Lag-Augmented Vector Autoregression (LA-VAR) model proposed by Toda and Yamamoto (1995) and Dolado and Lütkepohl (1996). The sample period spans from 1 July 2020 to 14 November 2022, with a minimum window size set at 20% of the sample. Models incorporate four augmented lags and a trend. Dashed lines represent the critical values of bootstrapped test statistics at the 90% and 95% significance levels. The results are robust to heteroskedasticity.*

### *4.7. Stablecoins*

Stablecoins are a type of cryptocurrency designed to maintain a stable value by pegging their value to a reserve of assets, such as fiat currencies, commodities, or other cryptocurrencies. These digital assets provide users with the benefits of cryptocurrencies, such as fast and secure transactions, while minimizing the price volatility typically associated with them. As stablecoin transaction volumes and user activity increase, the demand for processing these transactions on the Ethereum blockchain grows, leading to higher transaction fees. This is due to the limited throughput capacity of the Ethereum network, which can only process a certain number of transactions per second. As more users compete for this limited resource, they are willing to pay higher fees to ensure their transactions are processed in a timely manner. Consequently, higher stablecoin transaction volumes and user activity contribute to increased Ethereum transaction fees, reflecting the network's scarcity of processing capacity and the competitive nature of the market for blockchain transaction processing.

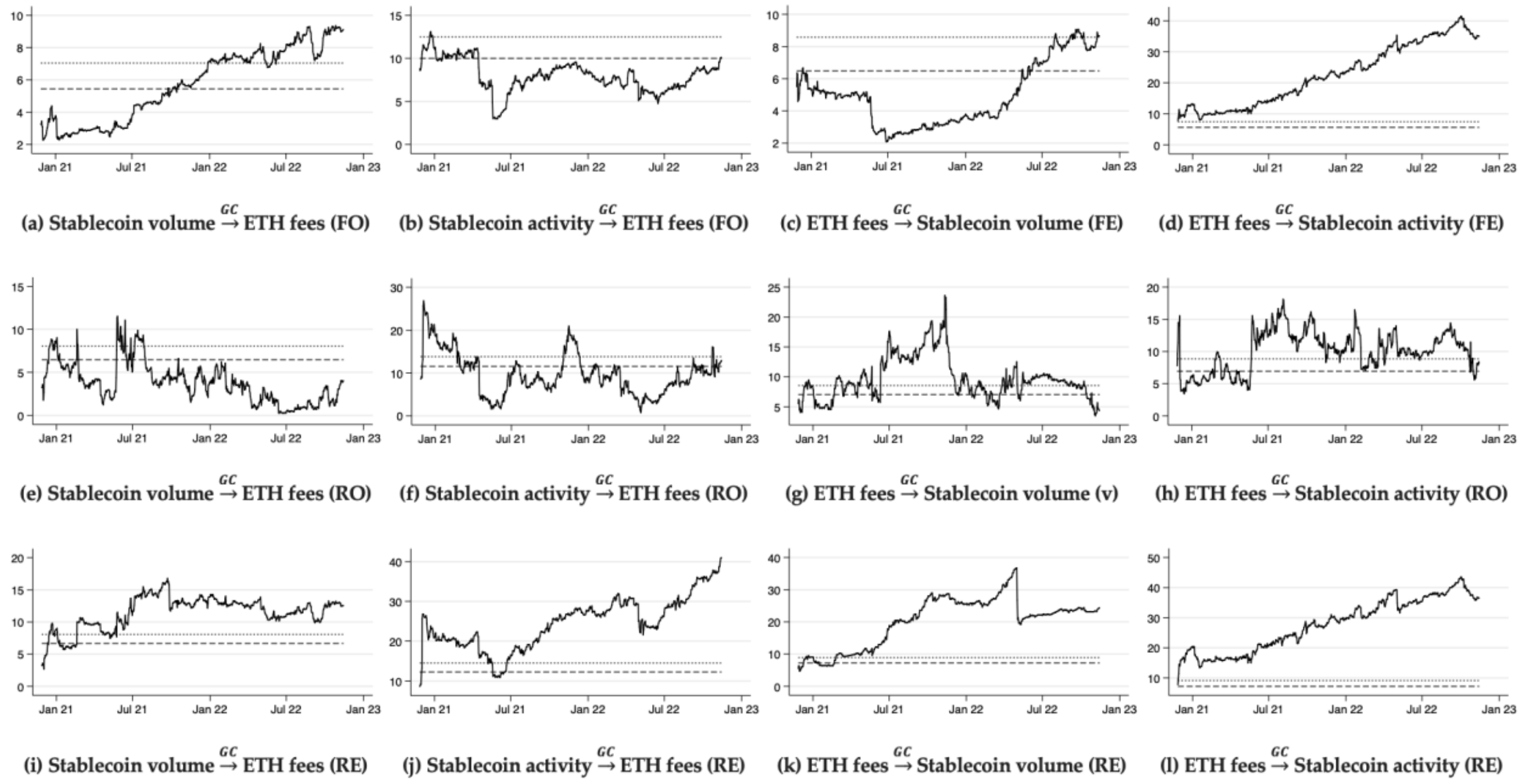

***Figure 9.*** *Time-varying Granger causality tests for stablecoin volume, activity, and mean fees in the Ethereum network. Notes: Figure 9 displays the bivariate results derived from forward expanding (FE), rolling (RO), and recursive evolving (RE) algorithms using the time-varying Granger causality model developed by Baum et al. (2021, 2022). The analysis employs the Lag-Augmented Vector Autoregression (LA-VAR) model proposed by Toda and Yamamoto (1995) and Dolado and Lütkepohl (1996). The sample period spans from 1 July 2020 to 14 November 2022, with a minimum window size set at 20% of the sample. Models incorporate four augmented lags and a trend. Dashed lines represent the critical values of bootstrapped test statistics at the 90% and 95% significance levels. The results are robust to heteroskedasticity.*

The time-varying Granger causality analysis between stablecoin transaction volume, user activity, and mean fees in the Ethereum network, as shown in Figure 8, reveals intriguing insights into the interdependencies between these factors. Our findings indicate that Ethereum transaction fees significantly Granger-caused stablecoin user activity (Figure 9, panels d,h,l) and transaction volume (Figure 9, panels g,k). Additionally, the recursive evolving (RE) algorithm detected a highly significant shift in Granger-causal directionality from Q2 2022 onwards, with stablecoin user activity (Figure 9, panel j) and transaction volume (Figure 9, panel i) Granger-causing Ethereum transaction fees. This feedback loop suggests a bidirectional causality that can be ascribed to the growing prominence and adoption of stablecoins within the cryptocurrency ecosystem. The increasing influence of stablecoins on the Ethereum network may have led to heightened transaction demand, consequently driving up Ethereum transaction fees. These findings highlight the complex relationships between stablecoin dynamics and the Ethereum network as they continue to shape the evolving cryptocurrency market landscape.

## 5. Discussion, Future Research, and Conclusions

Transaction costs represent a critical component of economic exchanges within blockchain ecosystems such as Ethereum, where they arise from actions like transferring the native cryptocurrency Ether or implementing smart contract operations. Within the Ethereum network, transaction costs (i.e., gas fees) remunerate the computational effort required to process transactions. A comprehensive grasp of these costs and their inherent dynamics is crucial, as transaction costs underpin all economic activities occurring within blockchain ecosystems like Ethereum. Moreover, these costs significantly influence the system's behavior and determine the feasibility of various economic endeavors. Therefore, an in-depth understanding of transaction costs in blockchain networks is indispensable for informed decision-making and the judicious exploitation of these advanced technological frameworks.

This study investigates the relationship between transaction fees within the Ethereum blockchain network and various economic subsystems that leverage the network, encompassing (a) Bridges; (b) CEXs; (c) DEXs; (d) MEV bots; (e) NFTs; and (f) Stablecoins. Through the application of a dynamic Granger causality analysis, the study unveils intricate, interconnected causal interdependencies between the transaction costs of the Ethereum network and its economic activities across these subsystems. The complexity of these relationships can likely be attributed to the direct influence of transaction costs on users' incentives to engage in economic transactions within the network.

Considering the analysis of the six discrete economic subsystems discussed in this paper, specific implications for each can be discerned:

- (a) Bridges: For bridges, the results reveal a bidirectional causality between the number of unique active wallets associated with bridge protocols and the mean transaction fees within the Ethereum network. The observed feedback loop potentially indicates a migration of users towards alternative blockchain infrastructures. Despite the considerable decrease in transaction fees over the analyzed duration, it underscores Ethereum's diminished competitiveness in comparison to other blockchain networks and layer-2 solutions. These insights highlight the role of transaction fees in influencing user migration trends and the ensuing need for judicious oversight.
- (b) CEXs: For Ethereum network stakeholders, the findings highlight the crucial role of CEX deposits and withdrawals in the fee network's operation. The strengthening, bidirectional Granger-causal relationship between Ethereum fees and CEX transaction volume is underpinned by a feedback loop. This suggests that increasing CEX transaction volume catalyzes demand for block space and transaction processing competition, resulting in higher gas fees. This, in turn, influences trading and transferring costs on CEXs, prompting users to pursue higher-value transactions, thereby reinforcing the causal nexus. Market participants may also monitor this interplay, capitalizing on arbitrage opportunities or market volatility, and perpetuating a self-reinforcing cycle of network congestion and escalating fees. Our findings contribute to the literature on centralized exchanges and decentralized blockchain networks (Ante et al. 2021a; Ante 2020; Aspris et al. 2021).
- (c) DEXs: The causal relationship between DEX volume, DEX users, and network fees illuminates the interplay among these three elements in the Ethereum network. The findings suggest that an increase in DEX volume causally influences higher fees, which subsequently have a significant causal influence on the DEX user counts. Over time, this relationship weakens, likely due to the diminished economic significance of the DeFi system (i.e., bubbles) (Maouchi et al. 2022; Wang et al. 2022). However, decreasing fees positively impacts the DEX user counts by rendering smaller trades economically viable again. Future scholarly inquiry is required to validate these postulations. For Ethereum network stakeholders, these findings underscore the need for DEXes to balance the trade-off between attracting more users and ensuring manageable fees, thus explaining why, e.g., Uniswap and SushiSwap also launched on other blockchain networks (Shen 2021) and continue to explore this option (Malwa 2023). Furthermore, DEXs need to consider the impact of fees on their user base when making fee-related decisions (i.e., network fees, not DEX-specific transaction fees). Additionally, the decline in fees' significance over time suggests that the impact of fees on users may differ depending on the economic context.

- (d) MEVs: A discernible causal linkage between Ethereum network fees and MEV volume/activity emerges during certain periods, signifying the intermittent importance of MEV within the Ethereum ecosystem. This phenomenon may be ascribed to elements such as the advent of alternative value-extraction approaches, alterations in searcher tactics, or adjustments in the Ethereum environment. Furthermore, market actors may have recalibrated their actions in response to the shifting interplay between MEV prospects and transaction fees, culminating in novel equilibrium points within the competitive arena. Subsequent investigations may consider delving into the potential ramifications of additional MEV market participants by employing a more exhaustive dataset, as the current findings, predicated on the activities of five eminent MEV bots, may not encompass the entirety of the MEV market landscape.[8]
- (e) NFTs: The analysis highlights a sophisticated causal interplay among NFT volume, NFT activity, and Ethereum network fees, where speculative bubbles may have significantly impacted relationships (Maouchi et al. 2022; Wang et al. 2022). Results show that fees causally influenced NFT activity and transaction volumes. This causal relationship can be interpreted as heightened fees acting as an entry barrier for users, discouraging (encouraging) NFT participation and resulting in reduced (increased) transaction volumes. Nonetheless, the evidence for the causal influence of NFT user activity or transaction volume on fees is not definitive, indicating a complex causal interplay. The lack of a consistent causal connection may be due to several confounding factors, such as network capacity, miner preferences, and overall transaction demand, with non-NFT-related transactions potentially exacerbating network congestion and raising fees independently of NFT activity or volume.
- (f) Stablecoins: Findings indicate that Ethereum transaction fees causally influenced stablecoin user activity and transaction volumes. Furthermore, evidence suggests a shift in causal directionality commencing from Q2 2022, wherein stablecoin user activity and transaction volume causally impacted transaction fees. This feedback mechanism infers a bidirectional causality, attributable to the burgeoning prominence and adoption of stablecoins within the cryptocurrency domain. The escalating influence of stablecoins on the Ethereum network precipitated heightened transaction demand, consequently leading to an increase in Ethereum transaction fees. These findings underscore the complex interdependencies between stablecoin dynamics and the Ethereum network as they collaboratively mold the dynamic cryptocurrency market landscape.

The findings of this study indicate that transaction fees serve as a significant causal determinant for nearly all investigated subsystems, and the activity and volume of numerous subsystems causally influence overall network fees. Nevertheless, it is essential to recognize the dynamic character of causality, which undergoes change over time and is plausibly attributable to various factors meriting additional exploration. By analyzing the activity and transaction volumes of each subsystem in conjunction with a time-varying Granger causality methodology, the causal spillover directions and temporal heterogeneity within the Ethereum network are discerned. This research presents a holistic evaluation of the interactions between Ethereum on-chain metrics across multiple economic subsystems, emphasizing causal spillover effects and temporal heterogeneity. The assessment of temporal variability uncovers a dynamic pattern of bidirectional causality between fees and primary economic markets, illustrating the evolving nature of these interactions across distinct timeframes.

The results of this investigation further corroborate the prevailing literature's assertion that the recursive evolving algorithm for detecting time-varying Granger causality produces superior and economically defensible outcomes, in alignment with the findings of other studies employing the same methodology (Ren et al. 2023). Conversely, the forward algorithm demonstrates the lowest level of detection precision, indicative of its incapacity to capture the most persistent causal relationships within the designated sample duration.

Future research could delve into specific subsystems such as bridges or NFTs, employing micro-level or user-level analysis to unearth granular insights. Such an approach would enable a deeper understanding of the individual behaviors and decisions that drive market dynamics within the blockchain ecosystem. Additionally, qualitative research methods could be employed to explore user motivations directly. Qualitative research surrounding questions, such as why individual users switched blockchains and if transaction costs influenced their decision, could provide valuable context to our quantitative data, offering a richer, more nuanced understanding of user behavior and market trends. This qualitative inquiry would not only complement our findings but also uncover the human factors influencing the technological and economic dimensions of blockchain networks.

Moreover, our study, while extensive, encounters limitations in the methodology regarding the direction of causality. This limitation presents a significant opportunity for future research to dissect and understand the causal relationships more precisely. A detailed exploration of causality directions could reveal the extent to which transaction costs directly influence user behavior and market movements, or vice versa. Such an investigation would be instrumental in refining our understanding of the complex dynamics within blockchain networks. Understanding these causal pathways is crucial for developers, policymakers, and stakeholders within the blockchain ecosystem to make informed decisions and strategize effectively.

Additional future research avenues include cross-blockchain comparative studies to understand the competitive landscape, the impact of layer-2 solutions on network efficiency, and the optimization of transaction processing algorithms. Behavioral economics and user behavior studies in response to fluctuating transaction costs, alongside the exploration of regulatory and policy implications, could provide deeper insights. Investigating the influence of technological innovations like sharding and rollups, conducting socio-economic impact studies for different user demographics, and interdisciplinary approaches combining computer science, economics, and sociology are crucial for a holistic understanding of blockchain ecosystems. These avenues not only extend current research but also foster interdisciplinary knowledge and practical applications in the blockchain domain.

In conclusion, it is important to acknowledge that the current paper does not provide an exhaustive analysis of the interactions between the examined variables. Additional research is necessary to ascertain the precise effects between fees and on-chain metrics, encompassing the number of unique active wallets and the extent of transaction volume. This study serves as a foundation for future inquiries to enhance the understanding of transaction cost dynamics within blockchain networks, such as Ethereum, and the implications for a range of stakeholders, including users, developers, and policymakers. By analyzing the intricate relationships between transaction fees and diverse economic systems on the Ethereum network, this research contributes to the burgeoning literature on blockchain networks, decentralized finance, and digital assets, ultimately facilitating more informed decision-making and efficacious utilization of these cutting-edge technologies.

## Notes

1. On 15 September 2022, the Ethereum blockchain underwent a substantial upgrade at block 15537393, commonly referred to as the Merge. This pivotal transition replaced the traditional proof-of-work (PoW) consensus mechanism with the more energy-efficient proof-of-stake (PoS) mechanism, where validators stake Ether in lieu of relying on hardware-based miners. Before the upgrade, the average block time experienced significant fluctuations due to network congestion. Post-merge, however, the block time has become more predictable and consistent, averaging approximately 12 s. This enhancement in block time can be ascribed to the accelerated and more efficient block processing facilitated by the PoS mechanism, as well as alterations to the transaction fee structure that have effectively mitigated congestion and augmented overall network efficiency.
2. Ethereum transaction fees are remitted in Ether; however, the associated 'gas' fees are denominated in Gwei, where one Gwei is equivalent to 0.000000001 Ether.
3. (Reijsbergen et al. 2021; Leonardos et al. 2021) determined that Ethereum Improvement Proposal (EIP) 1559 generally achieved its objectives, but suggested an alternative base fee adjustment rule employing variable learning rate mechanisms. Concurrently, Laurent et al. (Laurent et al. 2022; Azevedo Sousa et al. 2021) devised a novel Monte Carlo method to ascertain the minimum fee a user ought to pay for their transaction to be processed with a given probability within a specified timeframe. In contrast, Azevedo Sousa et al. (2021) found no evident correlation between Ethereum fee-related characteristics, such as user-defined gas and gas price, and the pending time of transactions. Lastly, Werner et al. (2020) introduced a gas price recommendation mechanism that amalgamates a deep-learning-based price forecasting model with an algorithm parameterized by a user-specific urgency value. This mechanism led to average cost savings exceeding 50% compared to existing recommendation mechanisms while incurring only a slight delay.

4. This study opts to utilize the USD value of the transaction fee as opposed to the Gwei value, as the former exhibits greater stability and is less susceptible to fluctuations. For instance, should the value of Ether experience rapid appreciation or depreciation, the corresponding Gwei value in USD would undergo swift alterations, which ultimately impacts users' focus and considerations. In addition, it can be assumed that economic players will want to use a stable currency such as the USD for their planning.

5. For comparison purposes, it is noteworthy to mention that PayPal's transaction fee structure involves a charge of 5% of the paid amount plus USD 0.05 (Khan and State 2020). Thus, only transactions exceeding USD 262 on PayPal would surpass the average transaction fee of USD 13.18 on the Ethereum network.

6. We validated this result using the KPSS and Augmented Dickey–Fuller test (Kwiatkowski et al. 1992).

7. See note 6 above.

8. In interpreting the results, it is vital to recognize the presence of survivorship bias within the underlying data. This bias arises from the consideration of only successful and valuable NFT collections (e.g., the five selected in this study), while numerous less successful or failed projects are excluded. These overlooked projects may constitute a larger market share and exhibit a distinct relationship with network fees. Consequently, the findings should not be generalized to the entire NFT market but rather pertain specifically to the upper echelon. This limitation extends to the analysis of MEV bots as well.